\documentclass[12pt,a4paper]{article}

\usepackage[pdftex]{graphicx}

\usepackage{amssymb, verbatim}
\usepackage{natbib, color, framed}
\usepackage{amsfonts}
\usepackage{amssymb}
\usepackage{amsmath}
\usepackage{graphicx}

\title{\fontsize{16}{12pt} \textbf{Exploiting new forms of data to study the private rented sector: strengths and limitations of a database of rental listings}}

\author{
Mark Livingston$^1$, Francesca Pannullo$^2$, \\
Adrian Bowman$^2$, Marian Scott$^2$, Nick Bailey$^1$ \\
~ \\
$^1$School of Political \& Social Sciences \\
The University of Glasgow \\ 
U.K.\\
~ \\
$^2$School of Mathematics \& Statistics \\
The University of Glasgow \\ 
U.K.
}

\begin{document}

\maketitle

{\small \noindent \textbf{Summary}
Reviews of official statistics for UK housing have noted that developments have not kept pace with real-world change, particularly the rapid growth of private renting. This paper examines the potential value of big data in this context. We report on the construction of a dataset from the on-line adverts of one national lettings agency, describing the content of the dataset and efforts to validate it against external sources. Focussing on one urban area, we illustrate how the dataset can shed new light on local changes. Lastly, we discuss the issues involved in making more routine use of this kind of data.
}

\


\newpage
\baselineskip=8mm


\section{Introduction}
\label{sec:introduction}

\subsection{The private renting data gap}

The biggest change in the UK housing system in recent decades has been the enormous growth in private renting, mainly at the expense of social renting but also, since the Global Financial Crash (GFC), at the expense of owner occupation. This is not a phenomenon unique to the UK \citep{Forrest:2015aa} but the pace of change here has been particularly rapid. The private rented sector has more than doubled in size over the last 20 years \citep{Scottish-Government:2018aa, Housing-Communities:2019aa}. While Government policy played a role in driving this change, this largely reflects a combination of market demand and supply factors \citep{Kemp:2015aa}. On the demand side, there have been declines in accessibility to mortgages due to the rises in insecure employment and to more cautious lending policies following the GFC as well as the loss of social housing. On the supply side, rising numbers of individuals have seen the sector as an investment opportunity as a result of the growth in the availability of mortgage products for landlords (the Buy-to-Let market). They have been further encouraged by the historically low rates of return on alternative investments. 

Various sources of official statistics can be drawn on to shed light on the private rental sector. These include a range of regular government surveys which can show the evolution of the tenure at national and regional levels and, in some cases, down to the local authority level. The Census provides much finer geographic detail but only once every ten years and with limited range of information -- none on rents, for example. A number of reviews of official statistics on the housing sector have noted the lack of data on private renting as a particular concern. Most recently, the review of statistics for housing and planning \citep{Office-for-Statistics-Regulation:2017aa} identified private renting as a key data gap. In particular, it found that a range of stakeholders wanted more robust information on rents, distinguishing new and existing lets, and available down to small area (sub-local authority) levels. In this respect, its findings closely echoed those for the earlier review of housing market statistics by the \citet{National-Statistician:2012aa}.

Certain characteristics of the sector make it particularly challenging to research, however. Ownership is very diffuse as the typical landlord owns only one or two properties, and their motivations or routes into landlordism are quite diverse \citep{Kemp:2011aa}. Unlike social renting, the sector is geographically dispersed while tenants tend to have a very high turnover \citep{Rugg:2018aa}. This is partly because the sector is attractive to those who want to remain mobile but also because private tenants in the UK enjoy minimal security of tenure \citep{Bailey:2007aa}. Not only does this make it easy for landlords to terminate tenancies, it makes it difficult for tenants to enforce their rights so moving is frequently the only realistic solution for tenants where they encounter poor management or substandard property conditions.

An additional motivation for obtaining better statistics on the sector is that it is home to a growing proportion of more vulnerable households, particularly working age households on low incomes or with children \citep{Kemp:2011aa}. This growth comes at the same time as `austerity' measures have been reducing the value of welfare benefits and restricting eligibility for benefits with a severe impact on the situation of low-income households \citep{Beatty:2017aa}, particularly those seeking private rented accommodation. Since 1988, tenants on low incomes have been able to apply for a means-tested benefit, Housing Benefit, which can cover up to 100\% of their rent. From the 1990s, policy changes began to restrict who could apply for the benefit and limit the amount they could claim \citep{Kemp:1995aa} but particularly severe reductions were imposed following the GFC \citep{Reeves:2016aa}. From April 2011 onwards, the maximum rent for which Housing Benefit could be claimed in the private sector was reduced from the median rent for the local area to the 30th centile rent for that size of property, while an absolute maximum level of rent was set for the whole country. Impacts were expected to be most severe in urban areas, especially London, where rents tend to be higher. These restrictions may be one of the factors driving lower income households out of the more central locations of UK cities \citep{Bailey:2018aa, Fransham:2019aa}. 

\subsection{Big data as the solution}

It is in this context that this paper explores the potential for big data to address at least some of our gaps in knowledge about private renting. Big data can mean many things but here we use the term to refer to `naturally-occurring data', i.e.\ data which may be useful for research or statistical purposes but which were not originally collected for those purposes \citep{Thakuriah:2017aa}. This can include data produced automatically by the normal functioning of digital systems (e.g.\ physical sensors for the environment or transport systems, or data from public administration or commercial business systems), volunteered or user-generated content (e.g.\ social media) and data generated by digital forms of surveillance (e.g.\ CCTV) \citep{Kitchin:2013aa}. The dataset we explore in this paper comes from the first category.

Big data and associated knowledge discovery technologies are seen as producing new possibilities for generating understanding, replacing domain expertise and theoretically-driven research with data science expertise and data-driven methods \citep{Miller:2010aa, Boyd:2012aa, Kitchin:2013aa}.  A number of challenges in exploiting big data for social research purposes have been identified. One is the technical challenge of assembling and analysing these data, most commonly referenced through the `three Vs' which highlight how these new forms of data differ from the conventional sources used by statisticians and researchers. The data may be large in \textit{Volume}, posing new challenges in data storage and handling. They may come in a \textit{Variety} of formats or structures, including unstructured formats, posing challenges for information retrieval and data integration . And they may be produced in a continuous stream (i.e.\ have \textit{Velocity}) as is the case with data from physical sensors, for example, posing challenges of managing and integrating newly-produced data with an established collection.  Proper exploitation will demand skills which are rarely taught as part of the traditional social science or statistical training \citep{Lazer:2009aa, Miller:2010aa}. A second challenge is that big data can raise important ethical concerns due to the scale of intrusion into individual lives which they make possible, particularly where multiple sources are combined, and given the lack of clarity over consent. \citet{Currie:2013aa}, however, makes the counter-point that there are ethical risks in not trying to make use of these novel forms of data. 

Neither of these points is critical in our case study, however. Scale or Volume is only sometimes the defining characteristic of these data \citep{Boyd:2012aa} and, indeed, in the case examined here, none of the `three Vs' is particularly relevant. While there are some technical steps in assembling the dataset as described below, these require only relatively routine data science expertise. Nor are ethical issues a significant concern since the data we focus on are about dwellings rather than people, and the information is from published advertisements. 

Instead, our case study highlights two other challenges associated with new forms of data. The first is the political-economic dimension: that big data lie predominantly under the control of private companies, in contrast to traditional, curated forms of social survey and Census data which are largely in public ownership. Ownership is fragmented and predominantly held by a small number of leading technology companies granted a dominant position by virtue of the first-mover advantages and economies of scale inherent in the digital economy \citep{Cukier:2013aa, Kitchin:2013aa}. New digital divides emerge as a result \citep{Boyd:2012aa, Halford:2017aa} with private firms occupying a dominant position relative to public and academic researchers. Divides can also arise within academia, where a select group are granted privileged access to data, or between academia and civil society. 

The second challenge we would highlight is around another `V' -- Validation or Veracity: the need to assess the quality and potential biases within any novel data source. The scale of big data does not, of course, remove issues associated with bias or incomplete coverage as few datasets cover the entire population of interest \citep{Boyd:2012aa}. The data emerge from systems designed for other purposes while the processes by which the research versions of these datasets are generated are often obscure. They may have undergone significant processing involving numerous subjective decisions before researchers get access to them. Big data must therefore be approached as, at best, partial and subjective representations of the social world \citep{Kitchin:2014aa}. Understanding dataset construction, coverage and bias are the essential first steps before any substantive analysis.  

Despite the limitations, there are nevertheless good reasons for engaging with these data. Their scale and timeliness offer the prospect of novel insights, particularly in relation to behaviours in the digital realm but also more generally. We therefore seek to respond to the invitation from \citet{Kitchin:2014aa} to critically engage with this opportunity. 

We are able to do so thanks to the UK's Economic and Social Research Council support for an initiative to enhance access to big data for academics and others through its Big Data Network 2014--19\footnote{https://esrc.ukri.org/research/our-research/big-data-network/; [Accessed 5 Feb 2019]}. The BDN was motivated in part by recognition of the first challenge identified above, namely the inequalities in access to these data. The Urban Big Data Centre (UBDC) is one of three originally funded by that initiative\footnote{Full disclosure: the authors of this paper were all Co-Investigators on the UBDC grant and/or employed through it.} and it identified work on private renting as a key priority, exploring various routes to open up access to novel data in that area. Of these, the most interesting has been the license agreement reached with one of the UK's major property listings services, \textit{Zoopla plc}.
The aim of this paper therefore is to describe the construction of a database of private rental lets from the records of this company, and to critically evaluate its strengths and limitations for analysis of the sector at national and local levels. We further illustrate its potential through modelling of spatiotemporal change in one market area, Glasgow. 

The structure of the rest of the paper is as follows. In Section~\ref{sec:data} we describe the process of constructing the dataset and outline the content. In Section~\ref{sec:analysis-validation} we then discuss the issues of data quality and means of attempting to validate the data by reference to other sources. In Section~\ref{sec:analysis} we report on our analysis of the data for one market area, Glasgow. This covers validation efforts and descriptives summaries of the data, before reporting on statistical models of rents in relation to location and time. To finish, we offer some conclusions in Section~\ref{sec:discussion} on the strengths and weaknesses of these particular data for the analysis of private renting in the UK's cities, and offer some more general observations on the potential use of big data for urban analysis in the future.


\section{Data and methods}
\label{sec:data}

\subsection{Constructing the database}

The database of rental listings was constructed by a team of social researchers and data scientists working at the Urban Big Data Centre (UBDC\footnote{www.ubdc.ac.uk}), based at the University of Glasgow. UBDC was funded by the Economic and Social Research Council (now part of UK Research and Innovation) as part of the national data infrastructure for social research. Its remit is to improve access for researchers to big data which can enhance our understanding of cities and urban processes. In this paper, we focus on forms of big data which emerge as the by-product of the administrative activities of private businesses but UBDC is interested in a much wider range of big data including those produced as by-product of public administration activities, or from on-line social activities, or by a range of sensing systems such as traffic or building sensors. 

The focus of UBDC efforts is determined by the domain expertise of the academics (social scientists and transport engineers) who lead its work. Private renting was identified as a priority area early on because it provided a perfect test case. First, it was a sector which had very high policy relevance, as discussed above. Furthermore it was a sector characterised by a number of challenges including lower property and management standards, poor affordability and high levels of insecurity for tenants \citep{Rugg:2018aa, Scottish-Government:2018aa}. Second, the availability of data on the sector had not kept pace with its growing importance in the housing system overall, as the comments of the statistical authorities noted above showed. 

To address these gaps, UBDC explored connections with the three largest rental listings firms which dominated the UK market at this time. Given its role as data infrastructure, the challenge here was not just to negotiate access for a specific research project or defined research team. Rather it was to secure access on terms which would permit the Centre to share data with a wide range of researchers (academic but also potentially others), with a wide range of possible research uses. Such a licence would have significant potential to overlap with a firm's business interests in relation to marketing their data and carried some commercial risks in terms of control over data assets. Overlap was minimised by restricting use to non-commercial research and risks were reduced through the use by UBDC of End User Licences. One firm was prepared to negotiate a licence on these terms, Zoopla plc, and a licence was agreed for an initial period, up to the end of 2018. The licence covered listings of properties for sale as well as for rent but we focus in this paper on the latter. 

The licence provided UBDC staff with access to Zoopla's current and historic UK data from 2009 via an Application Programming Interface (API) with a restricted volume of calls. Historic data cover listings which are closed or completed (the property has been withdrawn from the market, whether reflecting a successful sale or rental or not). Current listings are those still actively being marketed. Rental listings contain a number of fields. There is some basic information on the property, including address and postcode, type of property, size expressed in number of bedrooms, and rent sought. There are start and end dates for the advertisement period. Some additional information is provided in a text field and this may give further details on property characteristics. However this is unstructured and will not be consistent between properties. We do not therefore use that data here.

The process to retrieve historic data is a non-trivial one, that requires making a call to the API for every property on the \textit{Zoopla} property database, retrieving details for any listings for that property. The database represents most residential properties in the UK and so contains over 27 million properties. Given restrictions on volumes of calls, the process of retrieval of historic records from 2009 to 2016 took over 8 months. As 2009 data do not represent a full year, this paper will focus on 2010 to 2016 for which there were 3.8 million rental listings. 

\subsection{Data processing and cleaning}

Significant data processing and cleaning work was undertaken on the data downloaded from the API. Duplicate listings were identified (dates, postcode and rental value) and deleted. In addition, records were removed if they were missing critical data or had invalid values. Listings needed to have a valid start and end date, with the former preceding the latter. Valid postcodes were required to allow allocation to a local authority or other locations.  There are just under 2 million adverts after data cleaning, and these form $51$\% of the original adverts (Table~\ref{tab:numbers}). Estimates from the 2016 Family and Resource Survey (FRS) suggest there were $5.9$ million private rented households in 2016, which gives a broad indication of the ratio of Zoopla adverts to households. 

A number of factors suggest that many of the adverts deleted as part of the cleaning process are from the years before 2012. For example, missing dates are more prevalent in this period.
From 2012, a much higher proportion of records are complete suggesting that procedures to ensure data quality were tightened around that time. Since we are interested in assessing the potential for these data to inform our understanding of the sector and to be used more widely in social research and statistics, we focus on the period from 2012 onwards, where data quality is likely to be higher. Postcode centroids were used to generate geographical location.  This was done by merging with the May 2017 postcode file published by the Office for National Statistics. Where postcodes could not be matched these records were deleted as either incorrect or incomplete. A full discussion of data quality issues is provided by \citet{Livingston:2018aa}.

\begin{table}
\begin{center}


\begin{tabular}{lrr}
\hline
Reason & Number excluded & Percent  \\
\hline
Duplicated & 148,828 & 3.9\% \\
Missing dates & 1,701,009 & 44.5\% \\
Invalid  & 3,020 & 0.1\% \\
\hline
Total excluded & 1,852,857 & 48.5\% \\ 
Included & 1,967,359 & 51.5\% \\
\hline
\end{tabular}

\vspace{1em}

\begin{tabular}{lrrrrrrrr}
\hline
Year & Missing & 2010 & 2011 & 2012 & 2013 & 2014 & 2015 & 2016 \\
\hline
Duplicated & 32161 & 10850 & 16324 & 22022 & 15510 & 18936 & 14748 & 18277 \\
Invalid  & 79 & 7 &257 & 640 & 755 & 681 & 378 & 223 \\
\hline
\end{tabular}

\end{center}
\caption{Number of cases included and excluded.  Invalid cases arise from missing or invalid postcodes.}
\label{tab:numbers}
\end{table}

\subsection{Quality assessment and validation analysis}

The rental listings database has not emerged from a system of data collection designed specifically to capture either the whole of the sector, as the Census does, or a representative sample of it, as household surveys do. We have much less knowledge about the data generation process and we cannot therefore be as confident in the quality of the database. The key question, therefore, is the \textit{Validation} one: the extent to which it provides a representative picture of the sector.  The scale of big data is no compensation for bias. 

To explore data quality, Section~\ref{sec:analysis-validation} of the paper reports on a range of comparisons with existing data on the scale of the sector and on rent levels. On the former, we look at geographic variations by making comparison with the 2011 Census and we look at trends over time by making comparisons with data from a major household survey, the Family Resources Survey (FRS). On the latter, we compare rents in our database with those collected by national agencies with a remit to monitor private sector rents for the purposes of regulation welfare benefit payments. We also make comparisons with an experimental index of private rents published by the ONS. 

While validation is the most important challenge for researchers seeking to harness the potential of big data for social research,, we should be careful not to overstate the validity of the datasets against which we are `benchmarking' our novel collection. Household survey response rates have been declining for some time and, in the case of the FRS, are now around $54$\%. Even with the use of survey weights to bring the survey distribution of certain characteristics into line with the `known' distribution of these characteristics at national or regional levels, there is increasing potential for bias and incomplete coverage in these datasets.  Even the Census suffers from incomplete coverage. In 2011, an estimated $5$\% of households did not respond overall but, for those in private renting, the non-response rate was around $10$\% (ONS 2012 and associated on-line tables).

\subsection{Illustrative analysis}

The validation work is limited to comparisons with data at relatively high geographies, namely local authorities. The real potential of the listings database, however, lies in its ability to provide insights into changes at much finer spatial scales, and to track the development of the sector within cities and towns. To illustrate the potential here, we report the results of some initial modelling of the rental market in the Glasgow travel-to-work area. 


\section{Analysis for validation}
\label{sec:analysis-validation}

\subsection{Scale of the sector}

The first set of validation questions concerns how well the rental listings database, constructed from the records of one company, represents geographic variations in the private rented sector across Britain and its growth over time. There is an obvious risk that even a `national' operator might achieve greater market coverage in some geographic areas than others. There is also the concern that market coverage for one firm can change significantly over time, reflecting changes in competition from new entrants to the market, or mergers and acquisitions. The potential for rapid change in market share in an online business is high, with landlords and lettings agencies able to move business quickly if they perceive that one platform has established an advantage. This problem is reduced where properties are advertised on multiple sites but practices in cross-advertising can change rapidly as well.

To look at variations in geographic coverage, we compare the number of listings in each location with data from the 2011 Census as this is the only source capable of providing an accurate picture for smaller geographic areas. To do this effectively, we limit the Census data to those households living in the kinds of private rental property which would be advertised on the open market. We therefore exclude households in `tied' accommodation (rented with a job) and those living `rent free' (in a property owned by a family member or friend). 

\begin{figure}
\includegraphics[width = \textwidth]{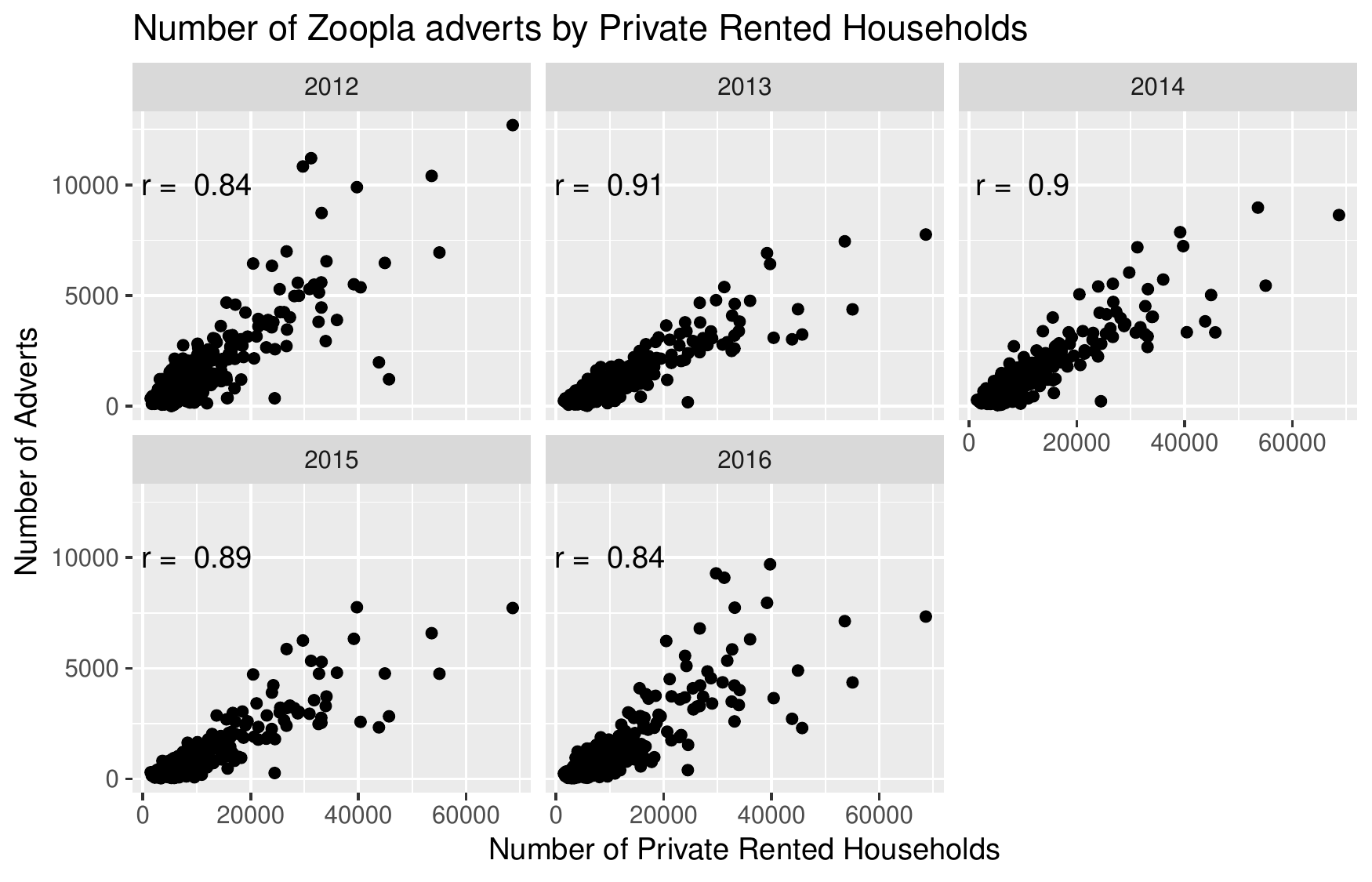}
\includegraphics[width = \textwidth]{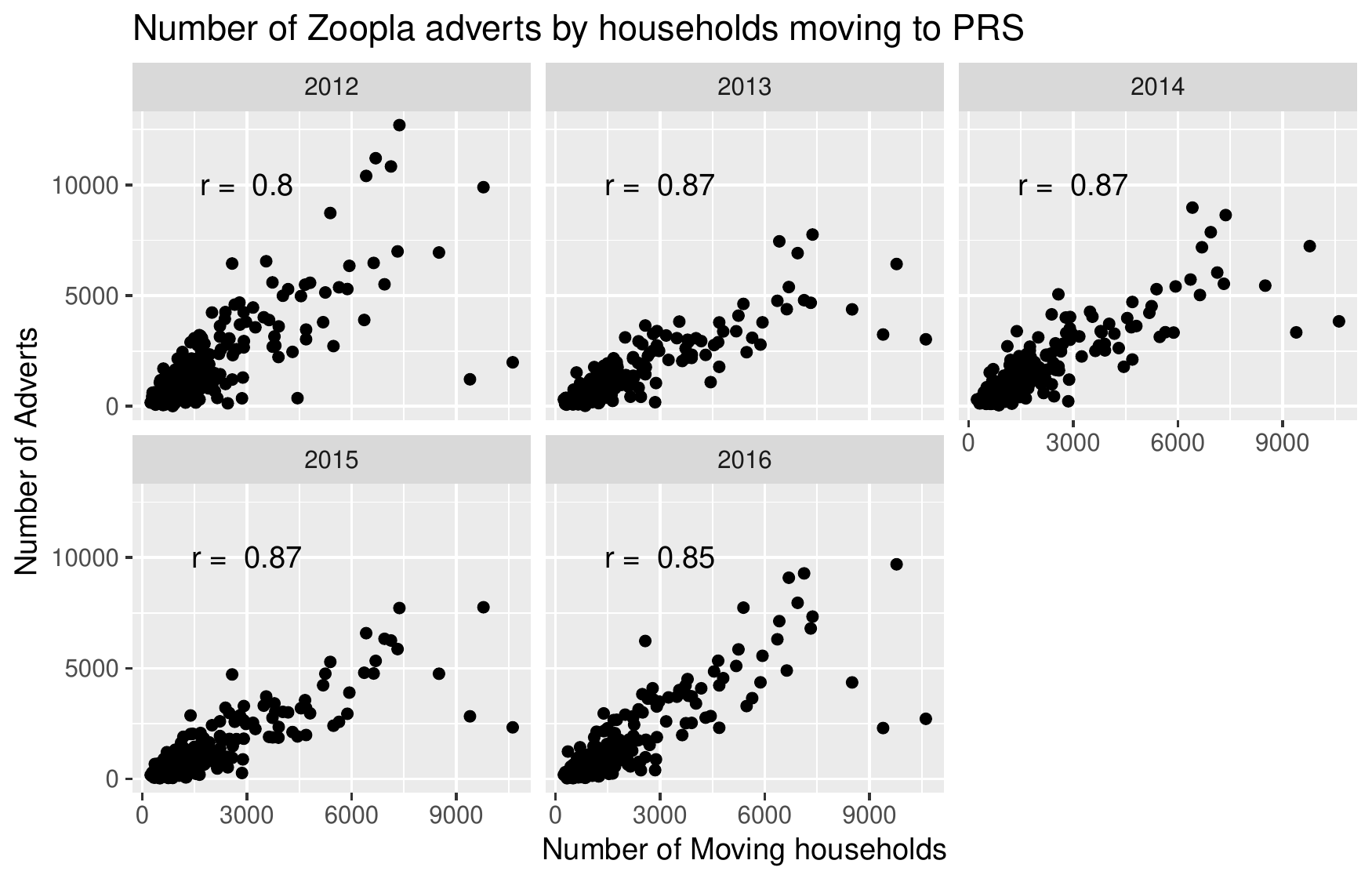}
\caption{The top set of panels show the number of listings (2012-2016) against the number of private rental households (Census 2011) in each local authority in Britain.  The bottom set of panels show the number of listings (2012-2016) against the turnover (Census 2011) for each local authority in Britain.}
\label{fig:adverts}
\end{figure}

Our database consists of listings for properties available for rent, and so measures the flow of lettings or the turnover of properties rather than the stock of households in the sector at a point in time. The Census provides both a measure of the stock of households and a measure of flow through data on the number of PRS households who had moved into their property in the previous year. We would expect the two Census measures to be highly correlated but, for completeness, we compare the number of rental listings in each British local authority to both measures. Figure~\ref{fig:adverts} shows the relationships for each year of data from the rental listings database. 

Overall, we find high correlations suggesting that our database has good geographic coverage. The highest correlations are found in 2013, with later correlations diminishing as time from the census date increases, as we would expect.  However, the correlations in 2012 are lower than in subsequent years. The reasons are not clear but appear to be do to some continuing problems with data quality around that time.

It is striking that the number of adverts has a higher correlation with the Census measure of housing stock than it does with the measure of the flow of new lets in the sector in most of the years we examine. We might expect that turnover rates or churn in the tenure would be higher in some areas than others, reflecting differences in the demographics of households or in the buoyancy of the rental market \citep{Bailey:2007aa}. If that were so, the flow of adverts ought to have a stronger relationship with the Census measure of turnover than with the measure of stock. On the other hand, turnover may be inherently a more 'noisy' measure, varying more from year to year, so that the stock of properties is a more reliable guide to the number of adverts we should expect in each area. Whatever the explanation, it is useful for our purposes that the number of adverts has a strong relationship with the known geographic spread of the tenure.

\begin{figure}
\centerline{\includegraphics[width = 0.7\textwidth]{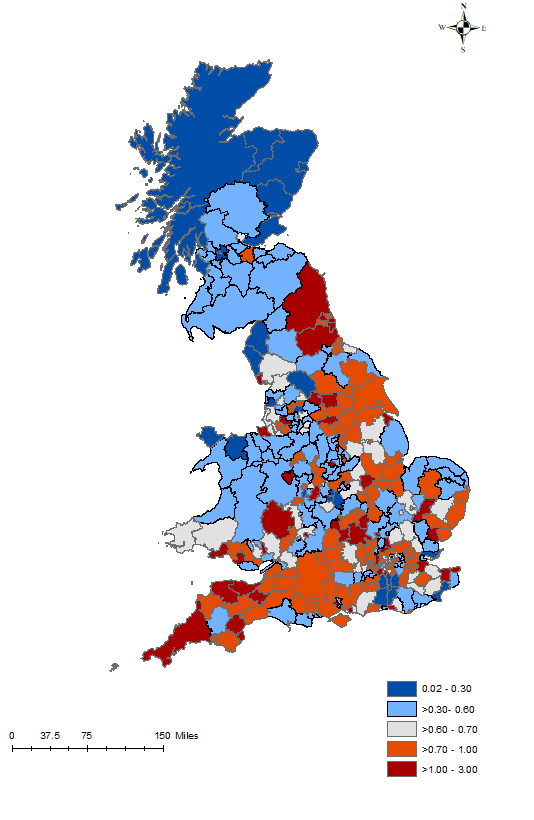}}
\caption{Ratio of adverts in 2013 to movers in census year 2011.  Sources: (1) Zoopla Property Group PLC 2018, (2) Zoopla Historic Data (UK to 2018), (3). Urban Big Data Centre.}
\label{fig:map}
\end{figure}

The data also allow us to estimate the approximate market coverage of the Zoopla database. The ratio of rental listings in 2012 to the Census estimate of the number of households moving in to a private tenancy in the previous year is approximately $0.95$. This suggests that approximately $95$\% of available lets are captured in this one database -- an astonishingly high coverage rate. If we look at geographic variations (Figure~\ref{fig:map}), we see a number of areas of the country where the ratio exceeds $1$. We can also identify areas, including much of Scotland, where the ratio is very low. The ratio of the number of adverts to Census stock suggests that one advert corresponds roughly to five private rental dwellings.

To look at changes in coverage over time, we need to make comparisons with national data from continuous sample surveys, in this case the FRS (DWP 2018). Table 2 compares the trends in the number of listings in our database with trends in survey estimates of both the stock of properties and the flow of lets (the number of entrants to a PRS dwelling in the last year). The Table shows the steady rise in the stock of PRS properties, up by $15$\% in this period, but also a slight fall in the turnover rate so that the flow of new entrants rose by just $5$\%. Trends in the number of listings do not follow these changes well. Not only did the number of listings in the rental database fall by $18$\% in this period, there was considerable volatility year-to-year. This suggests we would be unwise to use changes over time in the number of listings as a guide to changes in the stock of properties. 

\begin{table}
\begin{center}
\begin{tabular}{rrrrrrrr}
Year & \multicolumn{3}{c}{Numbers} & & \multicolumn{3}{c}{Index (2012 = 100)} \\
 & Listings & Stock & Flow & Turnover & Listings & Stock & Flow \\
  & ('000s)  & ('000s)  & ('000s) & rate & & & \\
2012 &	560	& 4,426	& 1,265	& 29\%	& 100.0	& 100.0	& 100.0 \\
2013 &	406 &  4,663	& 1,251	& 27\% 	& 72.5	& 105.4	& 98.9 \\
2014	 &	488	& 4,818	& 1,241	& 26\% 	& 87.2	& 108.9	& 98.1 \\
2015	 &	385	& 5,041	& 1,284	& 25\% 	& 68.8	& 113.9	& 101.5 \\
2016	 &	461	& 5,095	& 1,328	& 26\% 	& 82.3	& 115.1	& 105.0
\end{tabular}
\end{center}
\caption{Number of listings by year and stock of PRS properties.
1.	Zoopla Property Group PLC 2018. (2018). Zoopla Historic Data (UK to 2018). [data collection]. Urban Big Data Centre.
2.	Estimates from Family Resource Survey.}
\end{table}

\subsection{Rent levels}

The second set of validation questions concerns rent levels and whether the rental listings database provides an unbiased estimate of variations in rent levels across the country and over time. Although coverage is a little uneven geographically and has shifted markedly over time, it is possible that the database has retained a representative sample and can therefore provide unbiased estimates. On the other hand, providers may be specialised in certain sub-sectors of the market and so give a partial or biased representation.

To examine geographic variations in rents at local authority level, we make comparisons with a database of PRS rents constructed by the Valuation Office Agency (VOA). The VOA is responsible for gathering evidence on PRS rents across England in order to set limits for Housing Benefits and Universal Credit (i.e. to identify the 30th centile rent for the local market area) (VOA 2019). The VOA publishes data for regions and local authorities in England, showing means and median rents (as well as upper/lower quartiles) for different sizes of property over a twelve-month period. 

There are several differences between the VOA database and the UBDC rental listings databases. First the VOA is trying to identify rents for the stock of occupied properties rather than those currently available for rent. It does this by asking landlords and lettings agents to submit details of currently-let properties.  ``Valuation Office Agency (VOA) Rent Officers depend on the goodwill and trust of landlords, letting agents and tenants who provide details of rent levels being paid in the private rented sector (PRS). Around half a million PRS records are provided voluntarily each year. Together they form a unique database representing rents paid in England''\footnote{https://www.gov.uk/government/publications/local-housing-allowance-and-statistics-on-private-rent-levels}. Evidence suggests that landlords will often raise rents when properties turn over but avoid increases while they are occupied so we might expect the rental listings database to report slightly higher values than the VOA. Second, the VOA excludes any service charges (e.g. for fuel or water) as these are not eligible for Housing Benefits, whereas it is possible that some rental listings may include such charges. Third, it excludes any properties currently let to people on Housing Benefits. This would tend to remove cheaper properties from the database. All three factors would tend to lead to the rental listings database to record higher rent levels than the VOA.

Although the VOA data represent a very substantial effort to provide local data on PRS rents, it should not be regarded as a definitive benchmark. As the VOA notes, its dataset ``has not been drawn from a statistically designed sample, [so] the statistics in this release should be considered as indicative only of the private rental market'' (VOA 2019, p.6). Nevertheless comparisons between the two data sources can be informative. 

\begin{figure}
\includegraphics[width = \textwidth]{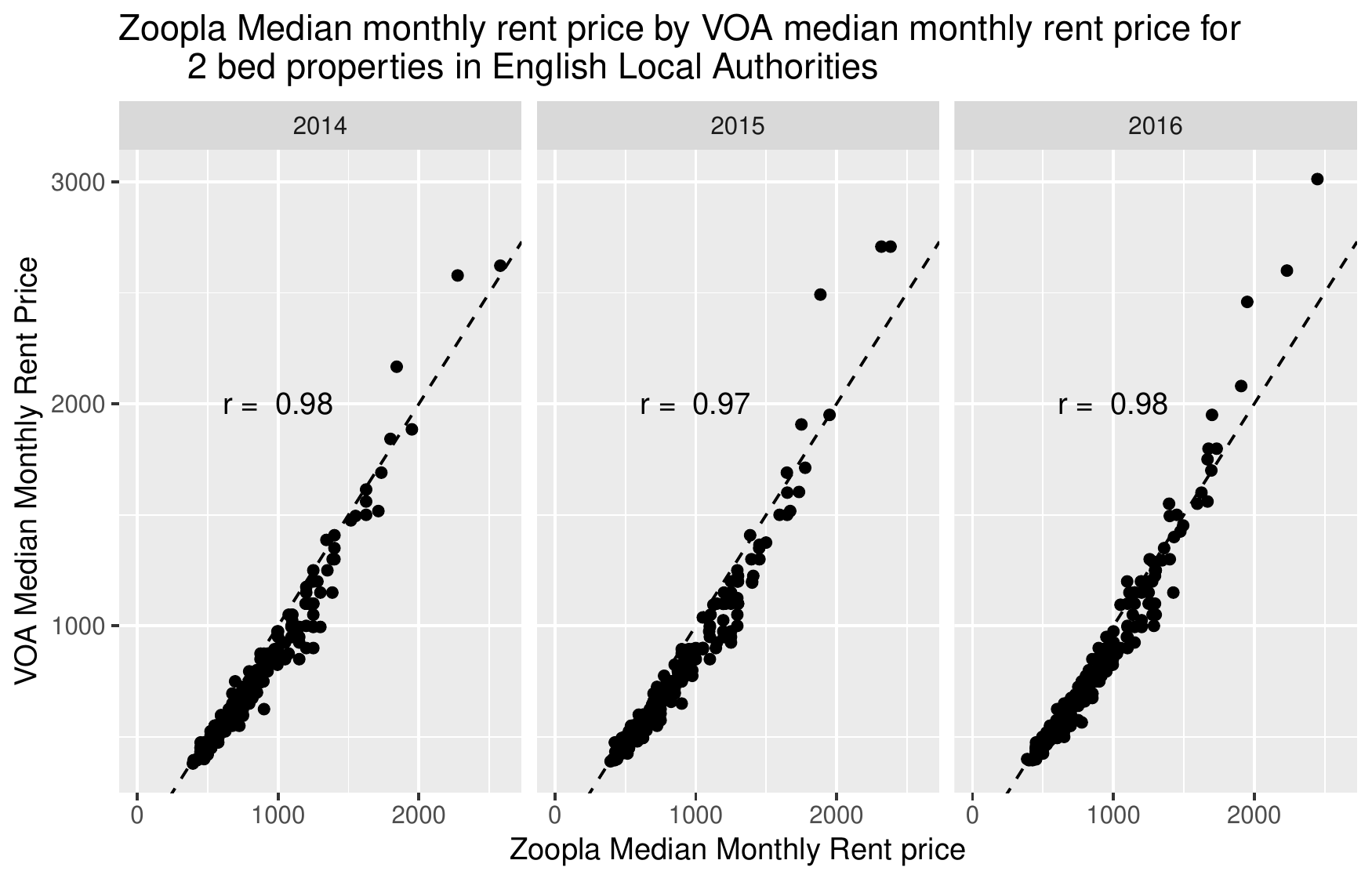}
\caption{VOA Median monthly private rent price and Zoopla median advertised rent price from 2014 to 2016.  The dashed line is the line of equality.  Sources: (1) Office for National Statistics, (2) Urban Big Data Centre, (3) Zoopla.}
\label{fig:rent}
\end{figure}

Figure~\ref{fig:rent} shows the relationship, for 2 bedroom properties, between the two estimates of median monthly private rent price and the Zoopla median prices for all local authorities in England for 2014 to 2016.  The scatterplots of the two sets of data across all three years in Figure 3 demonstrates very close correlations, with  R$^2$ values of $0.97$--$0.98$. The clustering around the line of equality shows that both sets of data are very closely aligned. This suggests that across England in different local authorities Zoopla rents are very close in value to the those estimated by the VOA, during a period when we know that the number of listings placed varies. 

\begin{figure}
\centerline{\includegraphics[width = 0.7\textwidth]{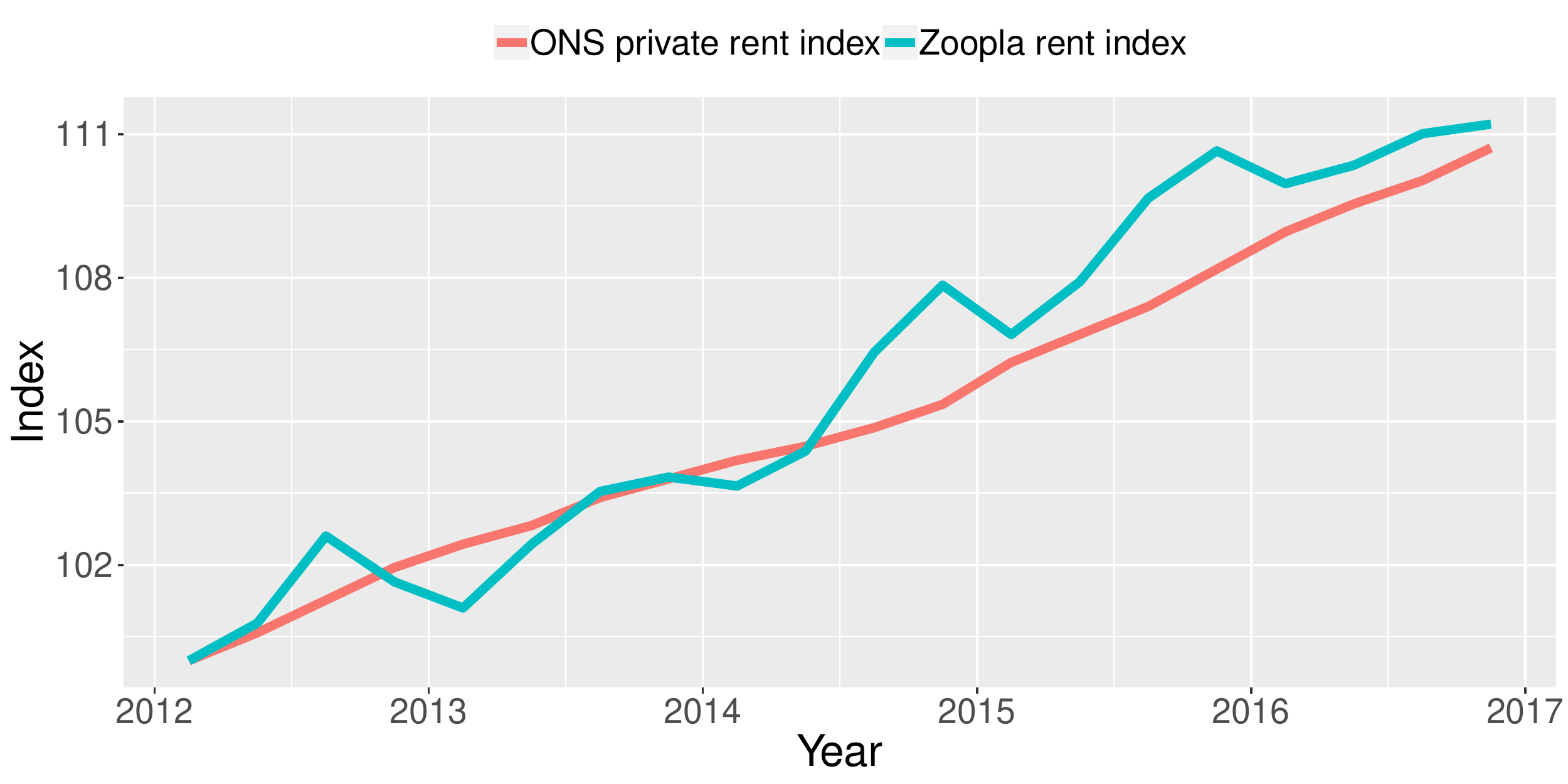}}
\caption{Rental indices from Zoopla data and ONS.  Sources: (1) Zoopla Property Group PLC 2018, (2) Zoopla Historic Data (UK to 2018), (3) Urban Big Data Centre}
\label{fig:index}
\end{figure}

The VOA (2019) stress that their data should not be used to construct indices of rents over time. To look at trends, we therefore compare changes in the rents in our rental database with ONS's Index of Private Housing Rental Prices (PHRPI). This is described as an `experimental index'. 

We estimate quarterly changes in rents from the rental listings database and compare with the Great Britain data from ONS (Figure~\ref{fig:index}). There is a fairly close relationship between the two estimates although, in the last two years, the rental listings database is suggesting somewhat greater growth in rents.


\section{Analysis of rental prices}
\label{sec:analysis}

On the basis of the reassurance offered by the validation reported in the previous section, the patterns of rental prices within one city, Glasgow, are now analysed, with particular interest in changes over time and space.  The precision of spatial locations and the linked information on property type provide a level of data resolution which is not possible to achieve from other data sources.   A number of different property types are represented in the \textit{Zoopla} data, including houses, semi-detached and terraced properties.  In order to provide a homogeneous dataset, analysis was restricted to flats, which form the dominant rental property type by far, with $11837$ adverts documented ($83$\% of the database).  Further homogeneity was achieved by restricting attention to properties which lie within a radius of $10$ miles from the city centre of Glasgow.  This includes areas which are part of the main conurbation and excludes isolated locations where different rental conditions may apply.  This reduced the dataset to $10626$ adverts.  The variables of interest are:
\begin{center}
\begin{tabular}{rp{4in}}
   \texttt{logprice} : & the advertised rental price in \pounds~per month, on a log scale; \\
   \texttt{beds} : & the number of bedrooms; \\
   \texttt{deprivation} : & a measure based on the proportion of residents in receipt of income benefit; \\
   \texttt{year} : & the year in which the advert appeared, expressed as a decimal to represent the exact date as a proportion through each year; \\
   \texttt{doy} : & the day of the year in which the advert appeared, to capture seasonal variations within the year; \\
   \texttt{location}  & the latitude and longitude  of the property.
\end{tabular}
\end{center}

\begin{figure}
\centerline{
   \includegraphics[width = 0.49\textwidth]{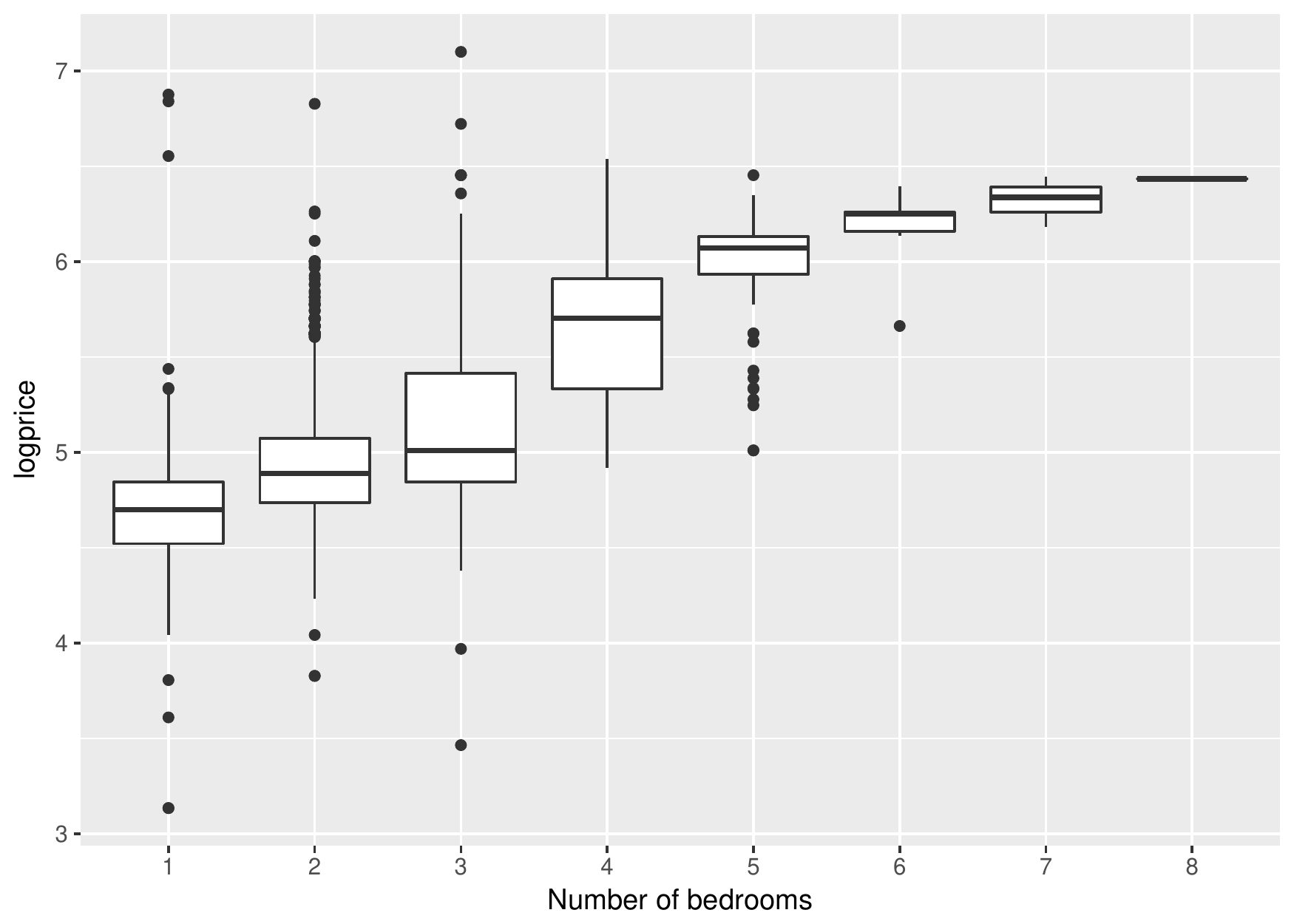}
   \includegraphics[width = 0.49\textwidth]{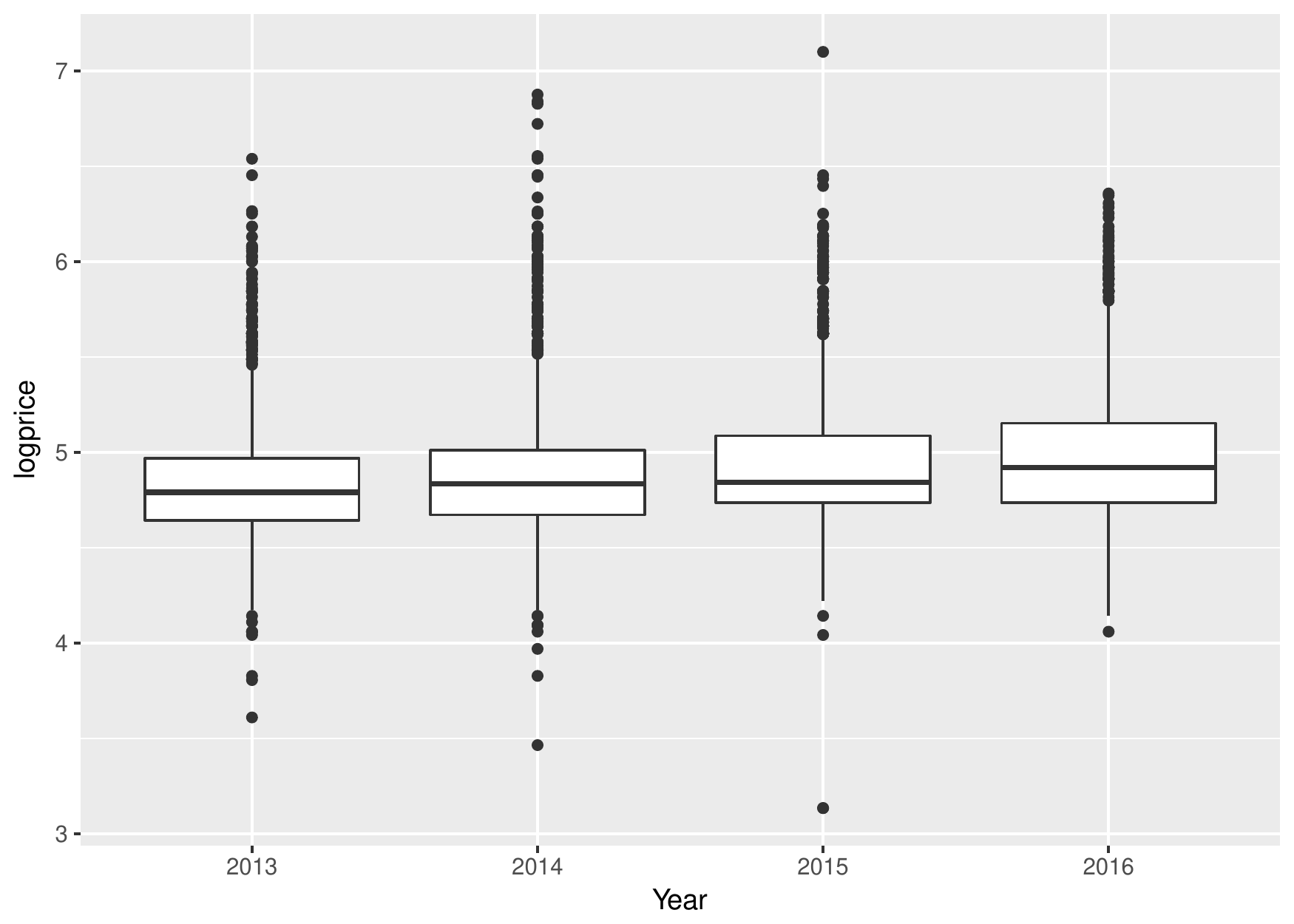}}
\centerline{
   \includegraphics[width = 0.49\textwidth]{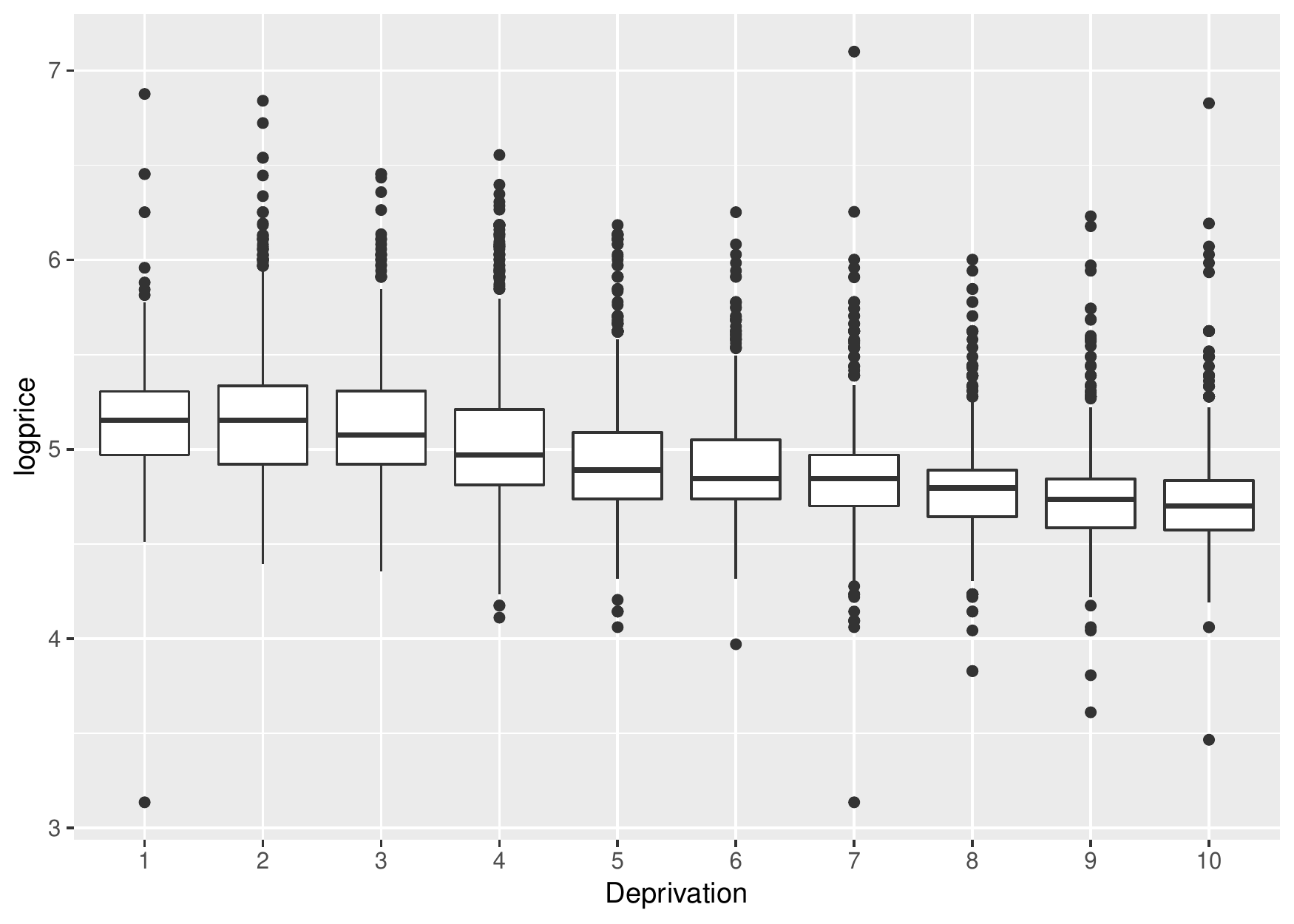}
   \includegraphics[width = 0.49\textwidth]{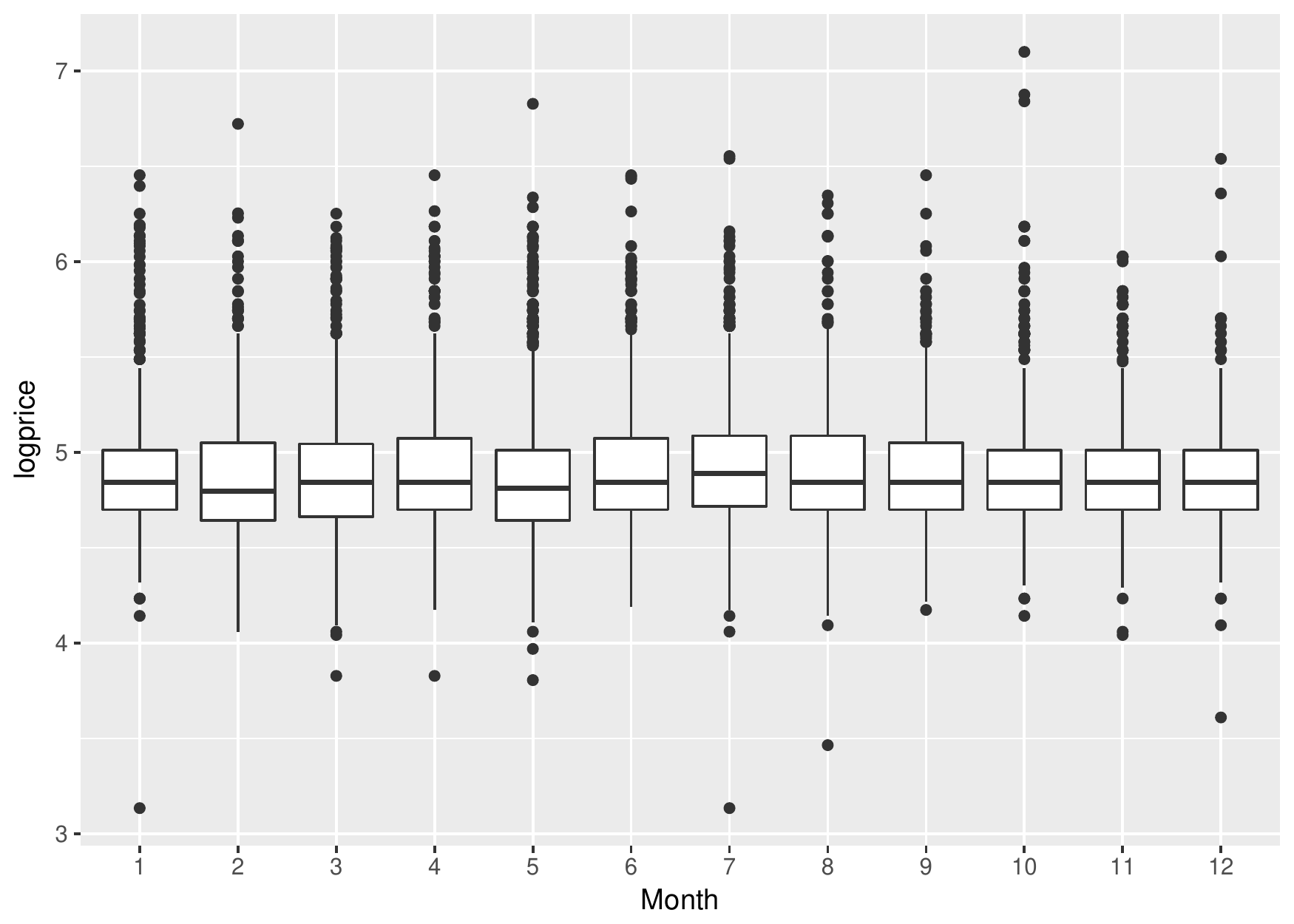}}
\centerline{
   \includegraphics[trim = {8cm, 4cm, 8cm, 5cm}, clip, width = 0.5\textwidth]{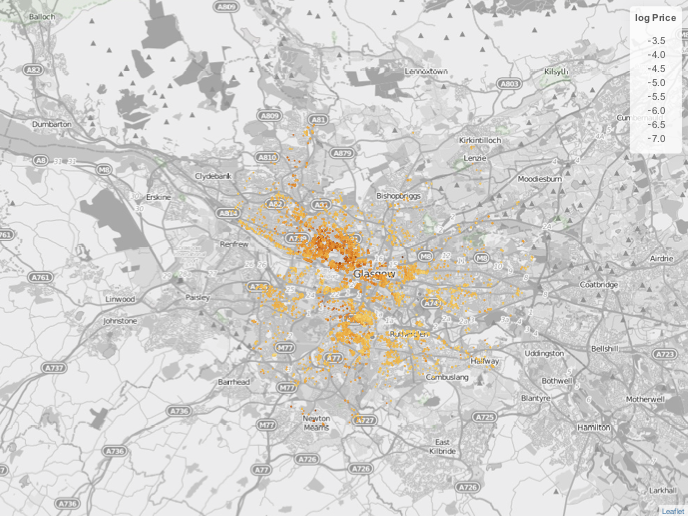}
   \includegraphics[width = 0.065\textwidth]{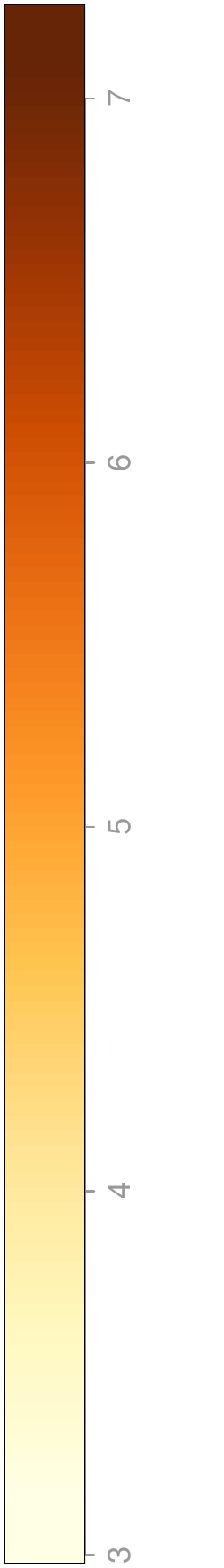}}
\caption{Boxplots of log price against four explanatory variables and a plot of the locations of rental properties colour coded by log price.}
\label{fig:boxplots}
\end{figure}

The marginal relationships between log price and the explanatory variables are shown in Figure~\ref{fig:boxplots}, with \texttt{doy} converted to months for clarity.  The number of bedrooms clearly has the strongest effect on price, and there are small effects of deprivation and year, but the day of year on which the advert first appeared seems to have little influence.  One of the most interesting features is the effect of spatial location.  This is of course constrained by strong geographical effects such as roads and rivers, and by estate boundaries.  The clustering of high value rental properties around the west end of the city is apparent.  The area in this district which is free of rental properties is where the University of Glasgow, Kelvingrove Museum and large parks are located.

The aim of a statistical model in this setting is to provide insight by teasing out the effects of individual variables and of their interactions.  A large dataset provides an opportunity to use models which are sufficiently flexible to track detailed effects, in contrast to standard models such as linear regression which impose a very rigid and potentially inappropriate structure on the relationships.  Additive models are therefore used here.  An excellent introduction to this form of modelling is provided by \citet{wood-2006-book}.  The model adopted here is:
\begin{eqnarray*}
\mbox{\texttt logprice} & \sim & \mu + m_b(\mbox{\texttt{beds}}) + m_a\mbox{\texttt{(deprivation}}) + m_y(\mbox{\texttt{year}}) + m_d(\mbox{\texttt{doy}}) + \\
                 &   &   m_l(\mbox{\texttt{location}}) + m_{by}(\mbox{\texttt{beds}}, \mbox{\texttt{year}}) + \\
                 &   &   m_{ay}(\mbox{\texttt{deprivation}}, \mbox{\texttt{year}}) + m_{ly}(\mbox{\texttt{location}}, \mbox{\texttt{year}}) + \varepsilon ,
\label{eq:model}
\end{eqnarray*}
where $\mu$ represents the overall mean and the other components represent smooth terms in the referenced variables.  The three interaction terms involving combinations of variables, here allowing changes in the effects of each variable over time to be investigated, as this is of particular interest.  As \texttt{year} is on a continuous scale, it is not appropriate to fit an interaction with the within-year variable \texttt{doy}.  

There are many ways in which the component functions can be estimated.  Here b-spline bases are used, with smoothness penalties on the b-spline coefficients, in the approach known as p-splines, introduced by \citet{eilers-1996-statsci}.  Bases for two- and three-dimensional model components are constructed from the product of the marginal bases for each contributing variable.  Identifiability is ensured by constraining the estimates of each main effect to sum to $0$ over the observed values of the contributing variables, while the coefficients of the interaction terms are constrained to sum to $0$ across each dimension, for every combination of basis indices in the other dimensions.  The details of the construction and fitting of additive models, for p-spline and other approaches, is described in \citet{wood-2006-book}.  The level of smoothness applied to each term in model (\ref{eq:model}) was selected by minimising the Bayesian Information Criterion ({\sc bic}) proposed originally by \citet{schwarz1978annals}.  In the case of normal errors this reduces to $n \log(RSS/n) + k \log(n)$, where $RSS$ denotes the residual sum-of-squares, $n$ denotes sample size and $k$ denotes the complexity of the model, expressed here in terms of the number of degrees of freedom (df) \citep[see][]{wood-2006-book}.  With the rental sector data, minimisation over the smoothness of each main effect term and allowing the smoothness of interaction terms to be inherited from these, led to the following choices: \texttt{beds} (4.1), \texttt{year} (1.3), \texttt{deprivation} (2.4), \texttt{doy} (5.5), \texttt{location} (26.4).

\begin{figure}
\centerline{
   \includegraphics[width = \textwidth]{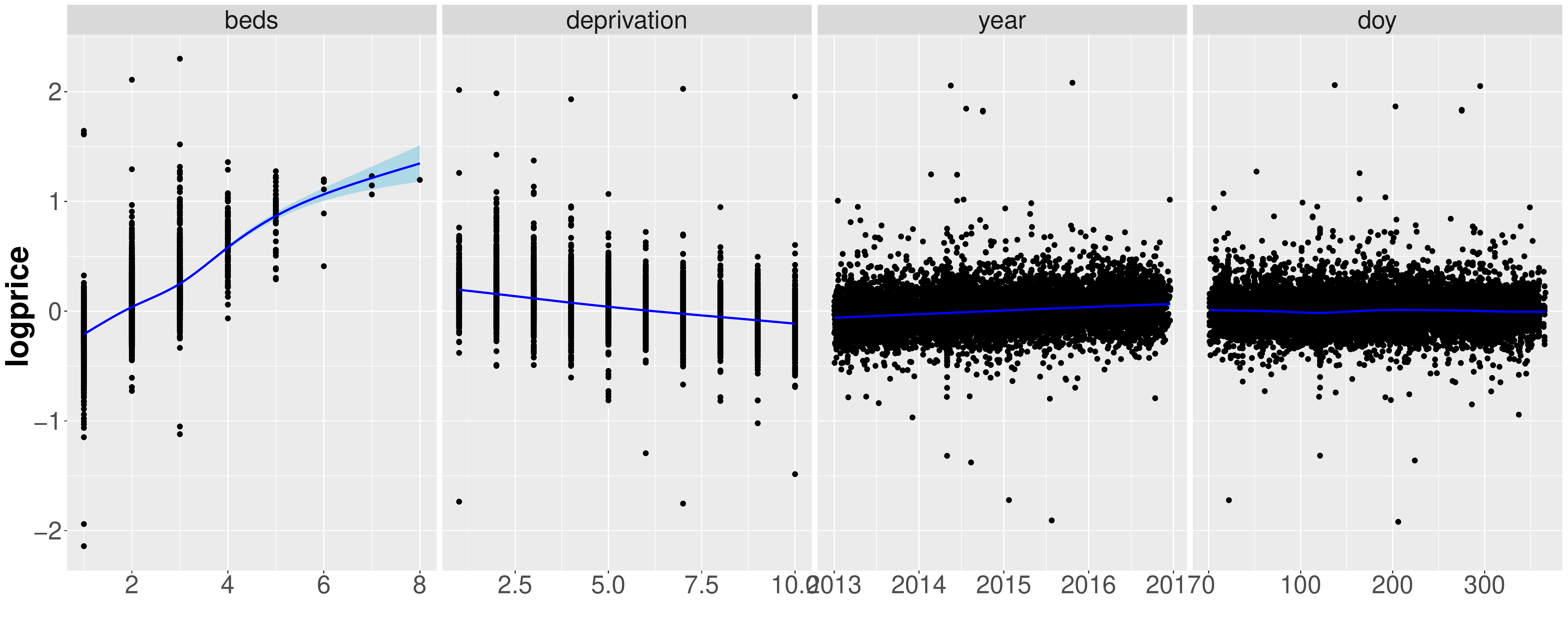}}
\centerline{
   \includegraphics[trim = {8cm, 4cm, 8cm, 5cm}, clip, width = 0.5\textwidth]{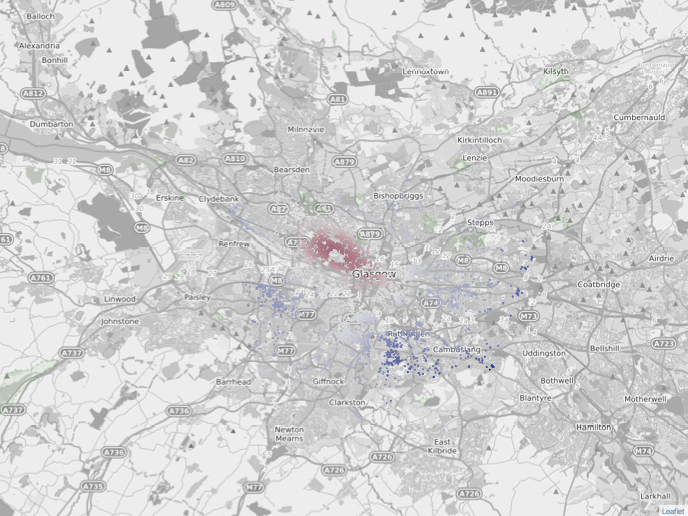}
   \includegraphics[width = 0.065\textwidth]{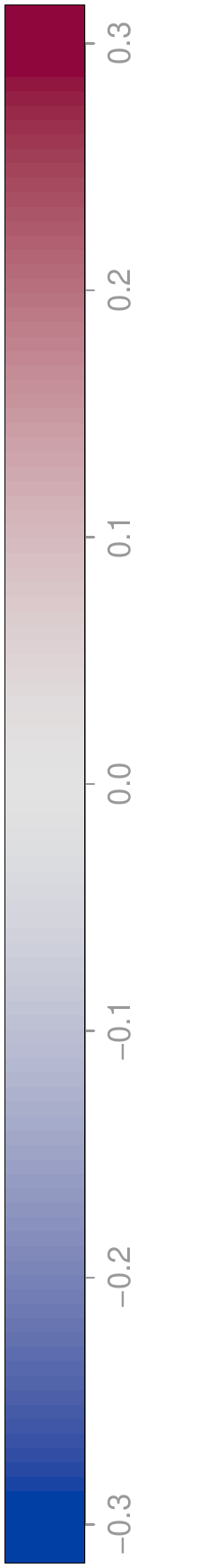}}
\caption{The main effect components of a fitted additive model.}
\label{fig:gam-main}
\end{figure}

The main effect terms of the fitted additive model are shown in Figure~\ref{fig:gam-main}.  These confirm the initial impressions from the plots of the raw data in Figure~\ref{fig:boxplots}.  For each additional bedroom over the range $1$-$5$, there is a multiplicative effect on price of approximately $\exp(0.25) = 1.28$.  There is a gentle decrease in price with deprivation, as a proxy for the characteristics of the residential area.  The main effect of year captures overall annual price rises of around $4$\%.  There is very little seasonal effect.  The effect of location is strong, reflecting the higher prices paid in the west end of the city, even after adjusting for the effects of the number of bedrooms and the deprivation characteristics of the area.  Although the spatial effect is modelled as a smooth surface across longitude and latitude, it is represented here through its evaluation at the observed location.  This allows the granularity of the spatial distribution of properties to be preserved.

There is a variety of ways in which the significance of terms in an additive model can be assessed, with \citet{wood-2006-book} again providing a helpful overview.  One simple strategy is to fit the model with a term of interest removed and repeatedly simulate from this reduced model, in a parametric bootstrap.  Re-fitting the full model to each simulated dataset provides the means of constructing the empirical distribution of the Wald statistic which quantifies the size of the coefficients associated with the model term of interest.  However, with a dataset of this size, significance tests are of little value as the large sample size generally leads to the rejection of simpler models.  This is the case with the rental sector data, so interest focusses on the size and interpretation of effects rather than on evidence of their presence.

\begin{figure}
\centerline{
   \includegraphics[width = 0.5\textwidth]{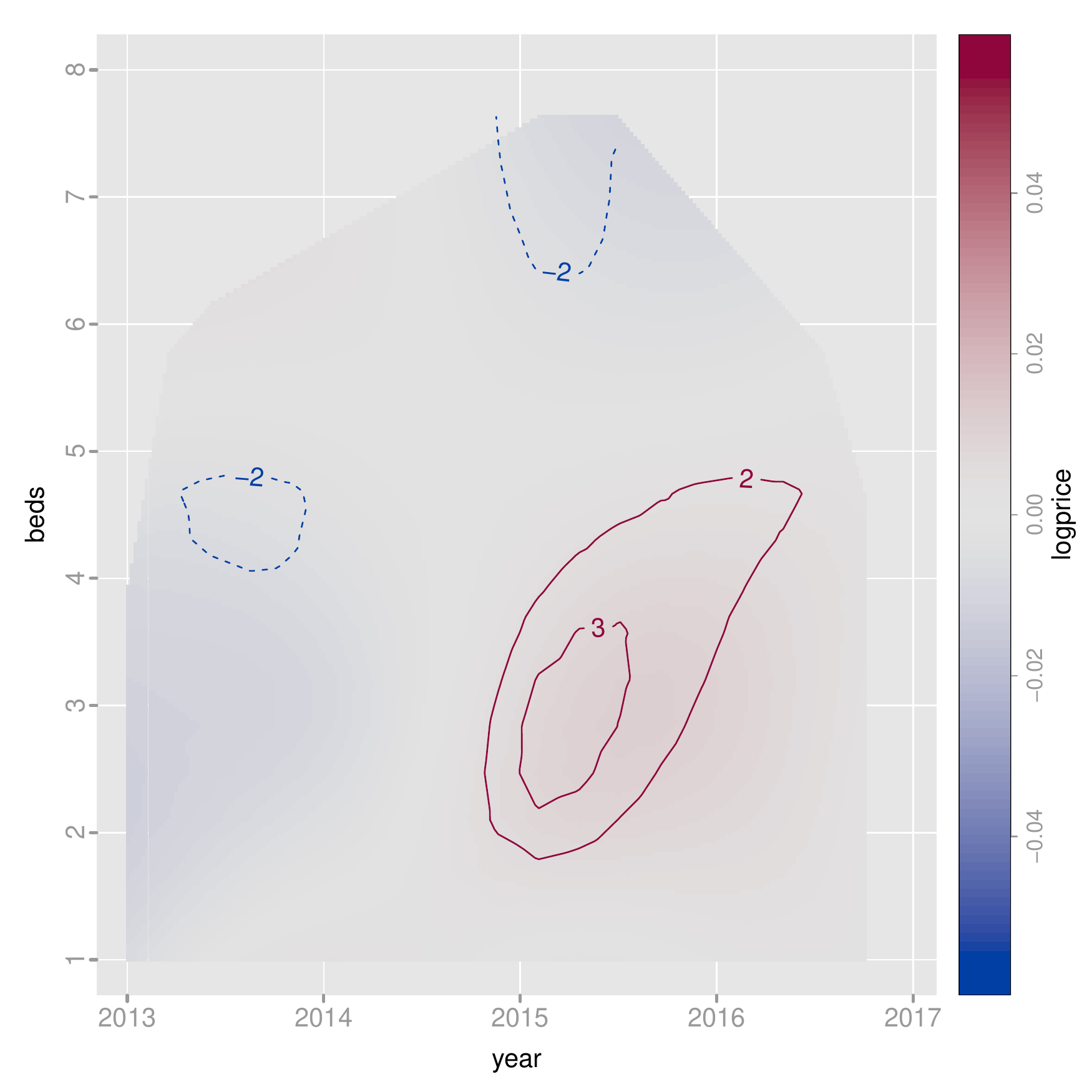}
   \includegraphics[width = 0.5\textwidth]{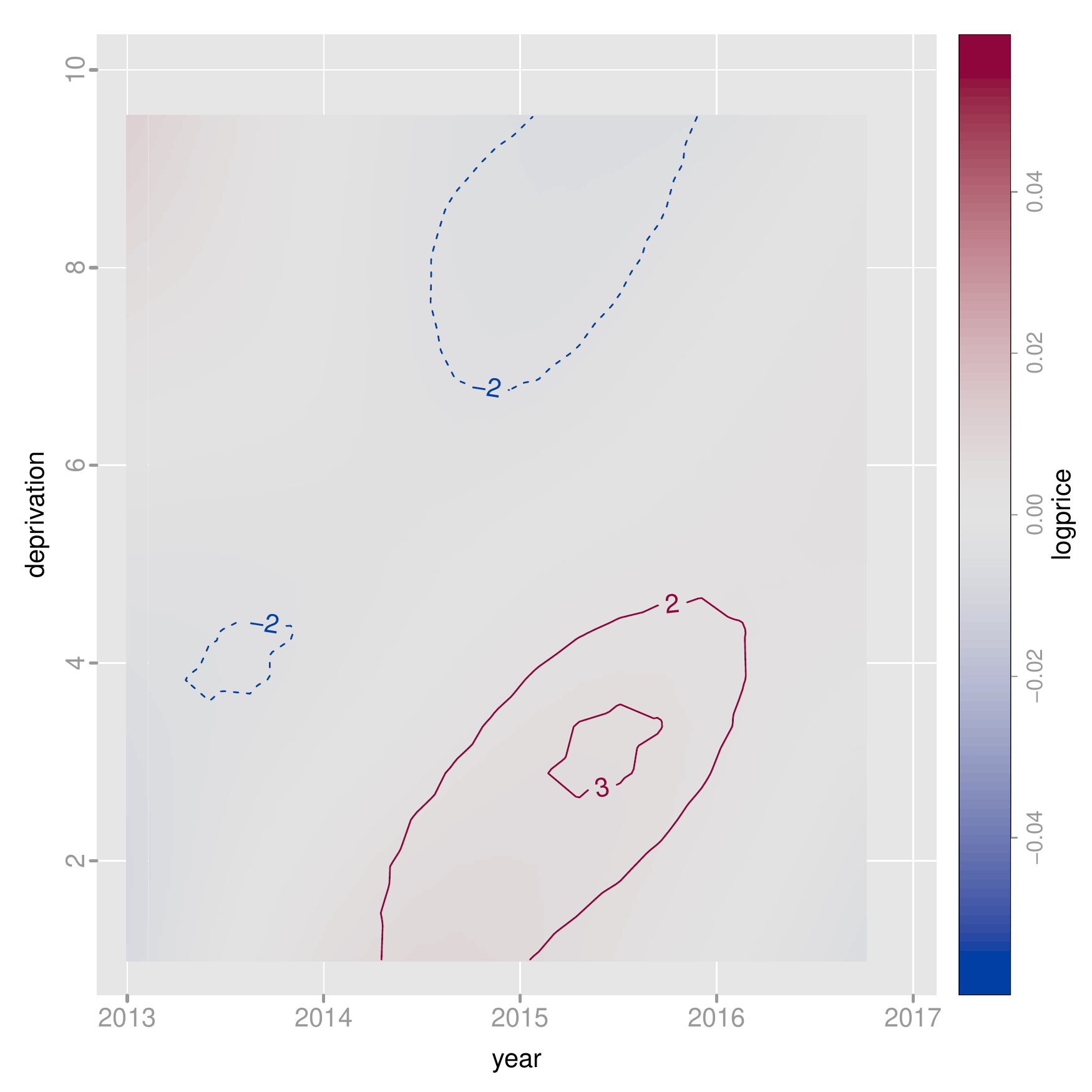}}
\centerline{~~~~~2013 \hfill 2014~~~~}
\centerline{
   \includegraphics[trim = {8cm, 4cm, 8cm, 5cm}, clip, width = 0.45\textwidth]{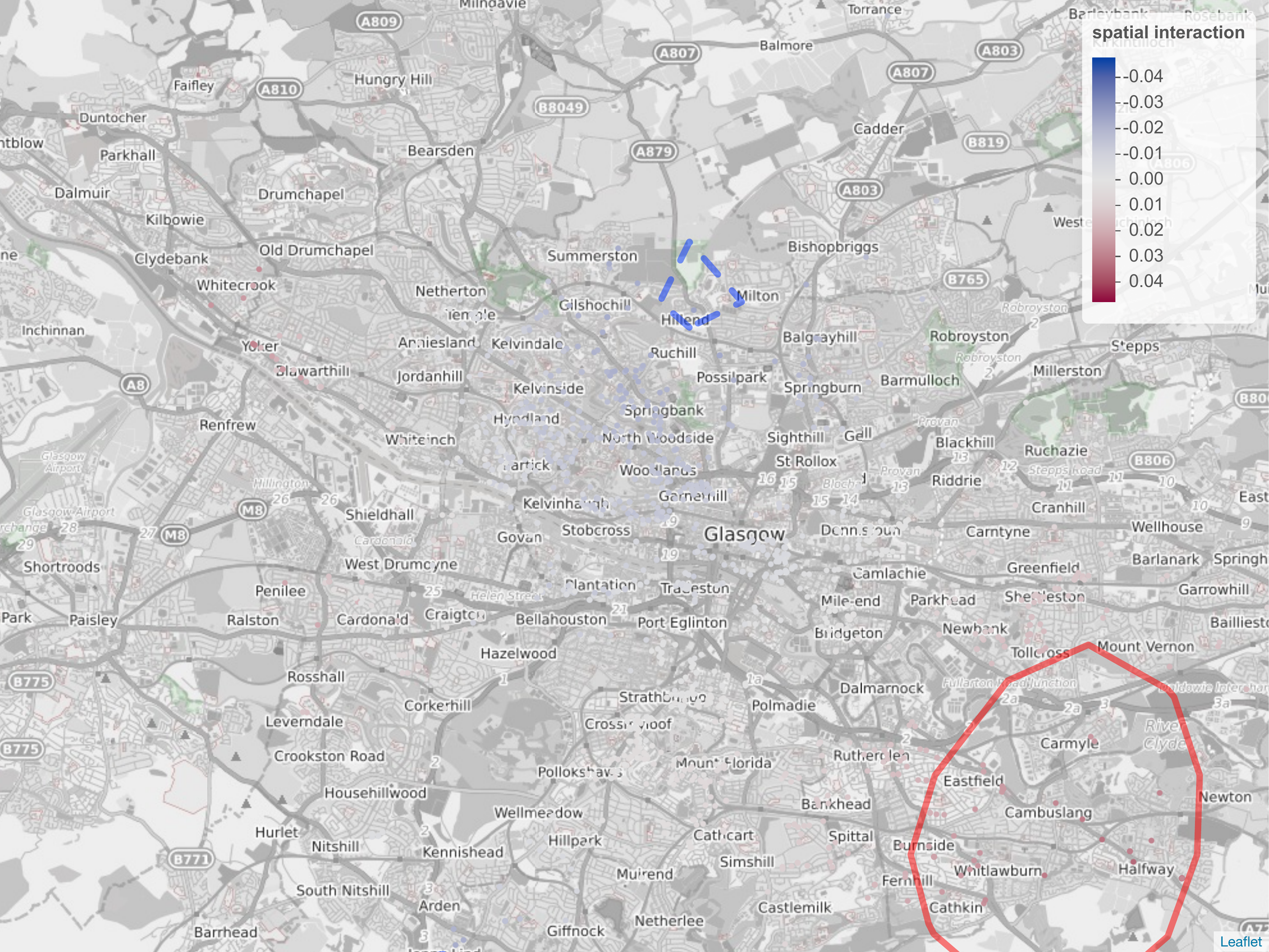}
   \includegraphics[trim = {8cm, 4cm, 8cm, 5cm}, clip, width = 0.45\textwidth]{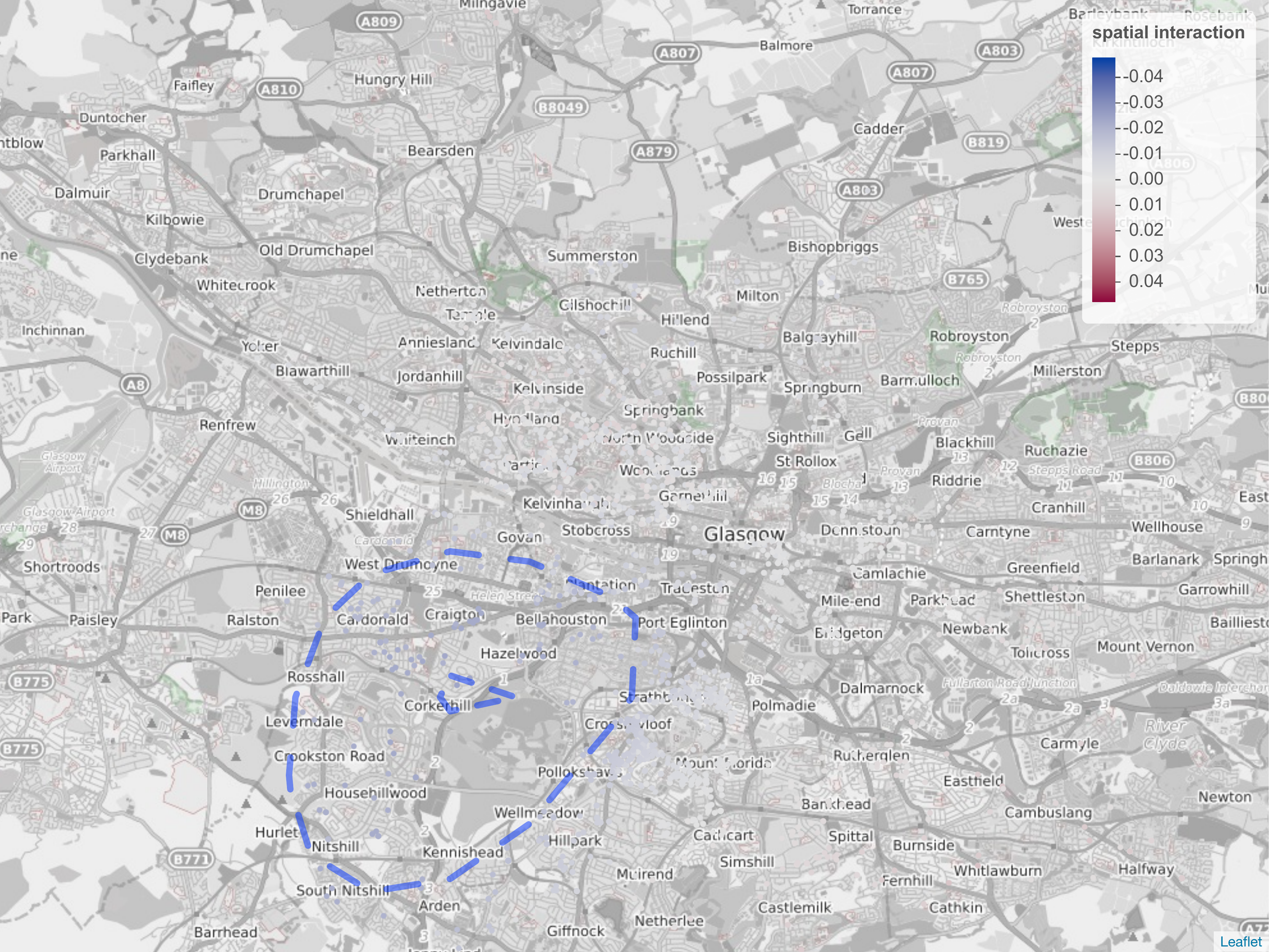}}
\centerline{~~~~~2015 \hfill 2016~~~~}
\centerline{
   \includegraphics[trim = {8cm, 4cm, 8cm, 5cm}, clip, width = 0.45\textwidth]{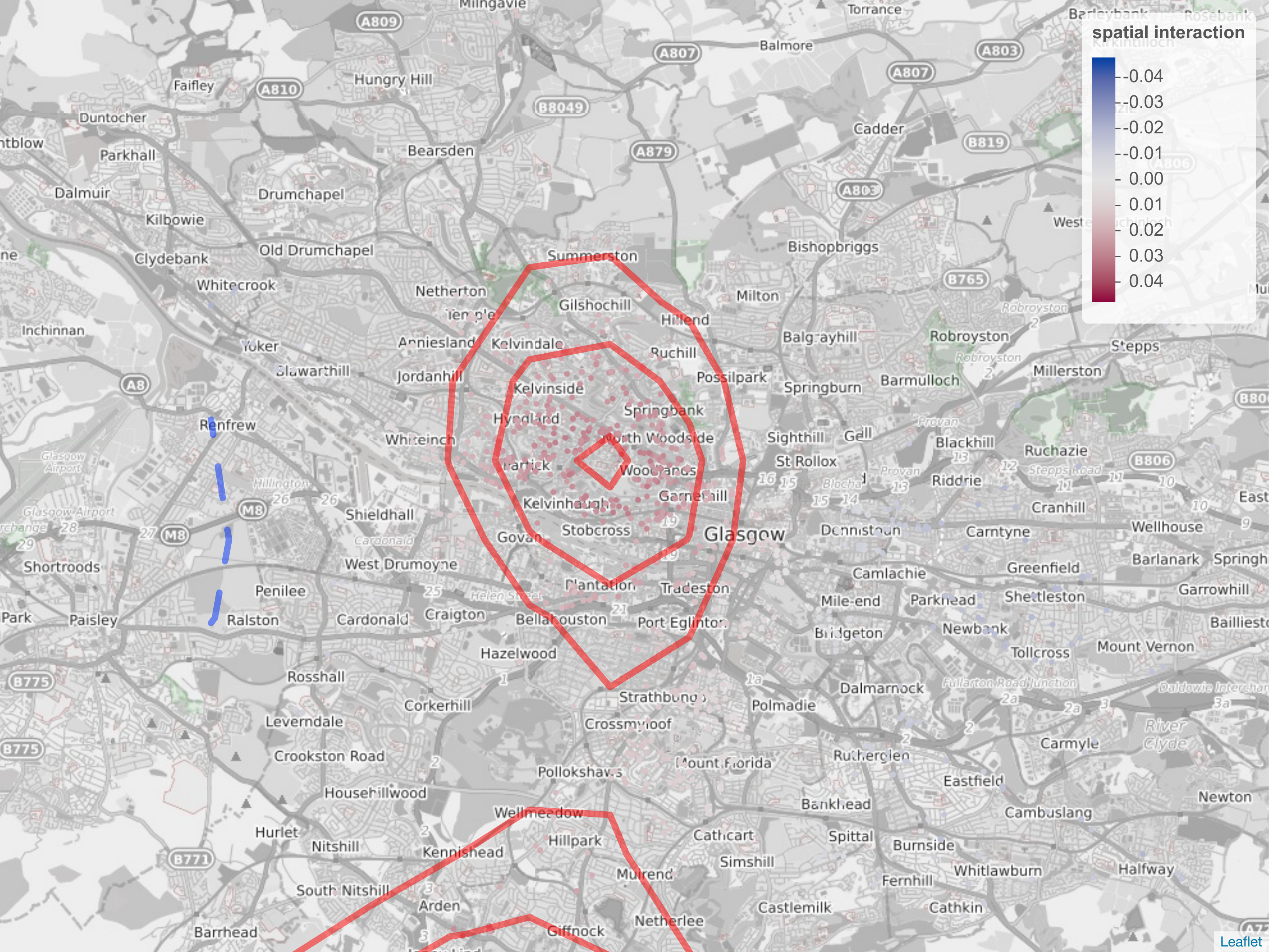}
   \includegraphics[trim = {8cm, 4cm, 8cm, 5cm}, clip, width = 0.45\textwidth]{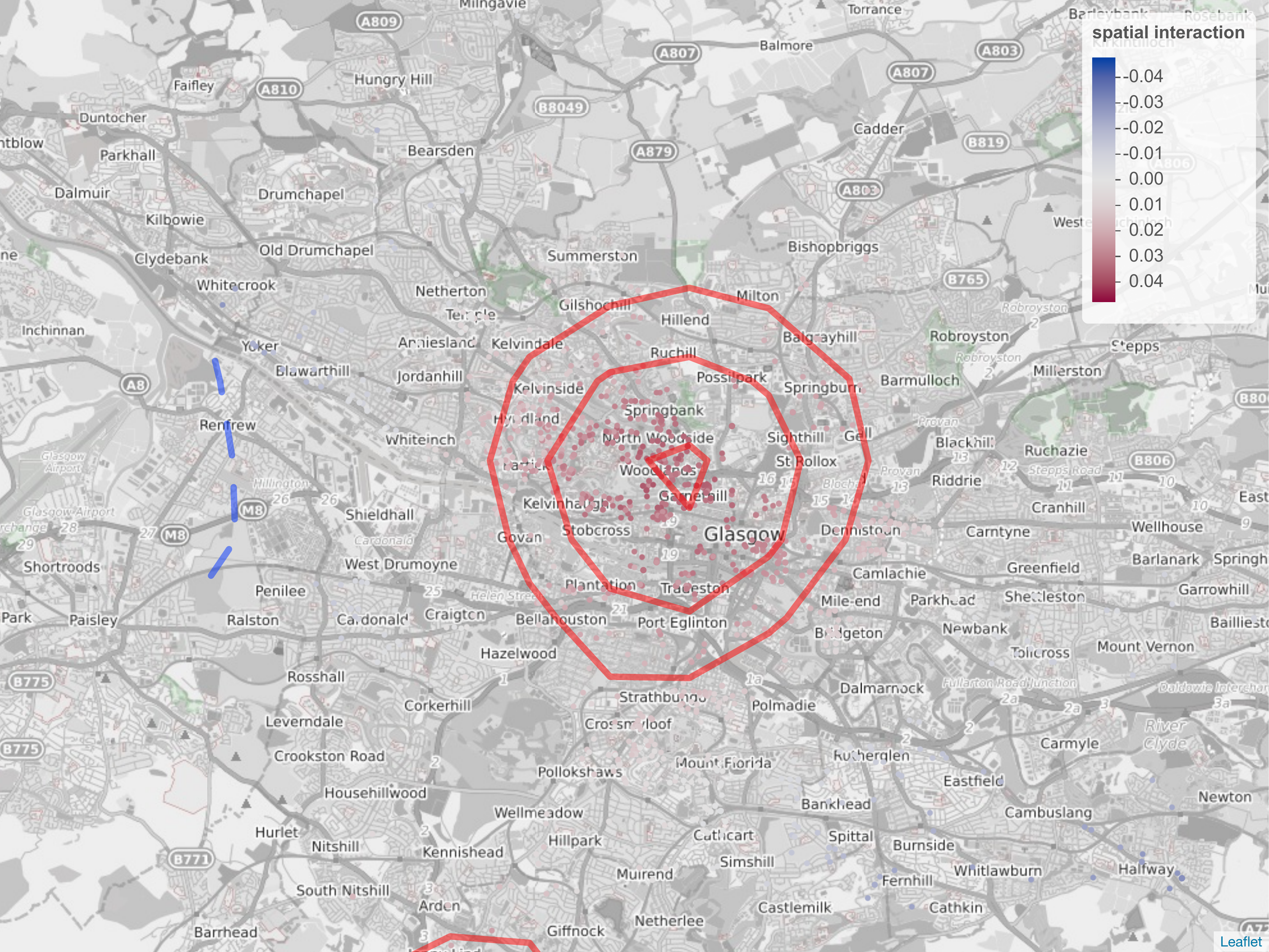}}
\caption{The top two panels show interaction terms for \texttt{beds} (left) and \texttt{deprivation} (right) with \texttt{year}.  The lower panels show the interaction terms for \texttt{location} and \texttt{year}.}
\label{fig:interaction}
\end{figure}

Interactions representing changing patterns over time are of particular interest and Figure~\ref{fig:interaction} illustrates this.  The surfaces shown here describe the adjustments to the main effects of Figure~\ref{fig:gam-main} which are required to capture the underlying patterns in the price variations.  These effects are displayed with a common colour scale but note that this has a much reduced range from the scale employed in the spatial component in Figure~\ref{fig:gam-main}, as befits second-order effects.  As an aid to the interpretation of these effects, contours are superimposed to indicate the regions where the estimated surface lies more than $2$ standard errors from $0$.  This is a device advocated by \citet{bowman-2009-applstat} and \citet{bowman-2019-jrssa} in the context of spatiotemporal models.  Here the contours show mild evidence of a slightly increased price for flats with modest numbers of bedrooms in 2014 and 2015 compared to the earlier two years, on top of the overall trend expressed in the main effect.  Similarly, there is a slight suggestion of an additional premium for flats in areas of lower deprivation in the later two years.

Indications of change in spatial patterns over time are of particular interest.  The lower plots of Figure~\ref{fig:interaction} display this, with little evidence of change in the early years followed by a small but marked premium in the west end in 2015 and 2016.  Again, this is an effect which should be interpreted as an adjustment to the main effect of location displayed in Figure~\ref{fig:gam-main}.

\section{Discussion}
\label{sec:discussion}

Despite potential shortcomings, big data do represent a positive opportunity for quantitative social science, informed by theory \citep{Kitchin:2014aa}. Their contribution is greatly strengthened by being part of a data landscape which includes curated social research datasets which provide a ground truth against which some elements can be tested, and a means potentially to correct for some biases.

Big data are often described as `naturally occurring' or `found' data but, as this paper makes clear, some forms of big data at least are better thought of as `achieved data'. To make these data accessible to researchers outside of the data-producing or data-owning organisation, is an extensive, time-consuming and complex process. This involves multiple skill sets, not just the obvious technical skills of data scientists to capture, wrangle and clean the data, but expertise in housing and the private rented sector specifically. licensing and copyright knowledge is also required to ensure the clear legal basis on which to use these types of data.

In this paper, we have demonstrated some of the issues associated with using one rental listings dataset as well as some of the strengths. Using the census and other nationally available data we have established that \textit{Zoopla} adverts are not stable in relation to stock over time, making them unreliable in this respect. Comparison with robust rental cost estimates are more reliable and while clearly these data do not represent the whole of the private rental sector, our analysis shows considerable stability in relation to rents, which show a high degree of correlation over time. We demonstrate that, despite concerns, it is possible to use these data to further our understanding of the private rented sector. The models from the Glasgow specific analysis show that these data can be used to explore subtle changes in rental prices.  The fine spatial and temporal granularity allows us to get beyond rent changes for the market as a whole and identify sub-areas within it with higher and lower rates of increase. This coincides with our informal knowledge of the rental market in the city, which suggests that the areas in which there are rising rental costs even after accounting for affluence have also seen significant growth of student numbers creating increasing pressure on this area's rental market. The use of these data allow us to more accurately identify rental sub markets and their location. More importantly, the ongoing availability of these data allow us to identify when markets are changing and where pressure may be occurring. These are analyses that are currently unavailable with other data sets.

We should be cautious about overstating the value of these kinds of big data at least. They have strengths but also significant limitations. In particular, where data are sourced from a subset of market providers, there can be rapid shifts in their coverage of the market so that reliance on these as a single source of knowledge would be very risky. Validation of the data against external sources is therefore essential and we show how that can be achieved here. To some extent, however, this undermines the very advantages of big data claimed by many enthusiasts: their timeliness. These data are at their strongest precisely when we can verify the picture they offer against systematically collected data. The most recent or timely data move beyond the period for which we have external validation, so we become unclear as to whether trends they show reflect change in the real world, change in market coverage or some combination of the two. Therefore these data require regular benchmarking if we are to remain confident of their usefulness.

There must also be some questions, therefore, about the longer-term viability of these data as a source to replace official statistics. To produce these data, we are wholly reliant on a licence agreement with the private company concerned but this is understandably time limited. When it is time to negotiate any extension to such a licence, it is unclear whether a given company will take the same view. On the one hand, if the data have proven useful and have come to be relied upon to supplement official statistics, the company would be in a more advantageous position in negotiations and could charge a higher fee. We might respond by seeking contracts through competitive tender but a change in provider would be a clear discontinuity in the data. On the other hand, the company's priorities may have changed, and it may see the supply of these data to a third party as undermining its new business model.


\section*{Acknowledgements}

The research reported in this paper was made possible by the Economic and Social Research Council's support for the Urban Big Data Centre (ES/L011921/1 and ES/S007105/1). Construction of the dataset involved the work of a large team of data scientists within UBDC. We are grateful to \textit{Zoopla plc} for their on-going support and collaboration in this work. None of these groups or organisations is responsible for any of the views expressed in this paper.

\bibliographystyle{Chicago}
\bibliography{rental-sector-paper}

\begin{thebibliography}{}

\bibitem[\protect\citeauthoryear{Bailey and Livingston}{Bailey and
  Livingston}{2007}]{Bailey:2007aa}
Bailey, N. and Livingston, M. (2007).
\newblock {\em Population turnover and area deprivation}.
\newblock Policy Press.

\bibitem[\protect\citeauthoryear{Bailey and Minton}{Bailey and
  Minton}{2018}]{Bailey:2018aa}
Bailey, N. and Minton, J. (2018).
\newblock The suburbanisation of poverty in british cities, 2004-16: extent,
  processes and nature.
\newblock {\em Urban Geography\/}~{\em 39\/}(6), 892--915.

\bibitem[\protect\citeauthoryear{Beatty and Fothergill}{Beatty and
  Fothergill}{2017}]{Beatty:2017aa}
Beatty, C. and Fothergill, S. (2017).
\newblock The impact on welfare and public finances of job loss in industrial
  britain.
\newblock {\em Regional Studies, Regional Science\/}~{\em 4\/}(1), 161--180.

\bibitem[\protect\citeauthoryear{Bowman}{Bowman}{2019}]{bowman-2019-jrssa}
Bowman, A.~W. (2019).
\newblock Graphics for uncertainty.
\newblock {\em Journal of the Royal Statistics Society, Series A, Statistics \&
  Society\/}~{\em to appear}.

\bibitem[\protect\citeauthoryear{Bowman, Giannitrapani, and
  Marian~Scott}{Bowman \textit{et~al.}}{2009}]{bowman-2009-applstat}
Bowman, A.~W., Giannitrapani, M., and Marian~Scott, E. (2009).
\newblock Spatiotemporal smoothing and sulphur dioxide trends over {E}urope.
\newblock {\em JRSS Series C (Applied Statistics)\/}~{\em 58\/}(5), 737--752.

\bibitem[\protect\citeauthoryear{Boyd and Crawford}{Boyd and
  Crawford}{2012}]{Boyd:2012aa}
Boyd, D. and Crawford, K. (2012).
\newblock Critical questions for big data: Provocations for a cultural,
  technological, and scholarly phenomenon.
\newblock {\em Information, communication \& society\/}~{\em 15\/}(5),
  662--679.

\bibitem[\protect\citeauthoryear{Cukier and Mayer-Schoenberger}{Cukier and
  Mayer-Schoenberger}{2013}]{Cukier:2013aa}
Cukier, K. and Mayer-Schoenberger, V. (2013).
\newblock The rise of big data: How it's changing the way we think about the
  world.
\newblock {\em Foreign Aff.\/}~{\em 92}, 28.

\bibitem[\protect\citeauthoryear{Currie}{Currie}{2013}]{Currie:2013aa}
Currie, J. (2013).
\newblock ``big data'' versus ``big brother'': on the appropriate use of
  large-scale data collections in pediatrics.
\newblock {\em Pediatrics\/}~{\em 131\/}(Suppl 2), S127.

\bibitem[\protect\citeauthoryear{Eilers and Marx}{Eilers and
  Marx}{1996}]{eilers-1996-statsci}
Eilers, P.~H. and Marx, B.~D. (1996).
\newblock Flexible smoothing with b-splines and penalties.
\newblock {\em Statistical science\/}~{\em 11}, 89--102.

\bibitem[\protect\citeauthoryear{Forrest and Hirayama}{Forrest and
  Hirayama}{2015}]{Forrest:2015aa}
Forrest, R. and Hirayama, Y. (2015).
\newblock The financialisation of the social project: Embedded liberalism,
  neoliberalism and home ownership.
\newblock {\em Urban Studies\/}~{\em 52\/}(2), 233--244.

\bibitem[\protect\citeauthoryear{Fransham}{Fransham}{2019}]{Fransham:2019aa}
Fransham, M. (2019).
\newblock Increasing evenness in the neighbourhood distribution of income
  poverty in england 2005--2014: Age differences and the influence of private
  rented housing.
\newblock {\em Environment and Planning A: Economy and Space\/}~{\em 51\/}(2),
  403--419.

\bibitem[\protect\citeauthoryear{Halford and Savage}{Halford and
  Savage}{2017}]{Halford:2017aa}
Halford, S. and Savage, M. (2017).
\newblock Speaking sociologically with big data: Symphonic social science and
  the future for big data research.
\newblock {\em Sociology\/}~{\em 51\/}(6), 1132--1148.

\bibitem[\protect\citeauthoryear{Kemp and McLaverty}{Kemp and
  McLaverty}{1995}]{Kemp:1995aa}
Kemp, P. and McLaverty, P. (1995).
\newblock {\em Private tenants and restrictions in rent for housing benefit}.
\newblock Centre for Housing Policy, University of York York.

\bibitem[\protect\citeauthoryear{Kemp}{Kemp}{2011}]{Kemp:2011aa}
Kemp, P.~A. (2011).
\newblock Low-income tenants in the private rental housing market.
\newblock {\em Housing Studies\/}~{\em 26\/}(7-8), 1019--1034.

\bibitem[\protect\citeauthoryear{Kemp}{Kemp}{2015}]{Kemp:2015aa}
Kemp, P.~A. (2015).
\newblock Private renting after the global financial crisis.
\newblock {\em Housing Studies\/}~{\em 30\/}(4), 601--620.

\bibitem[\protect\citeauthoryear{Kitchin}{Kitchin}{2013}]{Kitchin:2013aa}
Kitchin, R. (2013).
\newblock Big data and human geography: Opportunities, challenges and risks.
\newblock {\em Dialogues in human geography\/}~{\em 3\/}(3), 262--267.

\bibitem[\protect\citeauthoryear{Kitchin}{Kitchin}{2014}]{Kitchin:2014aa}
Kitchin, R. (2014).
\newblock Big data, new epistemologies and paradigm shifts.
\newblock {\em Big data \& society\/}~{\em 1\/}(1), 2053951714528481.

\bibitem[\protect\citeauthoryear{Lazer, Pentland, Adamic, Aral, Barab{\'a}si,
  Brewer, Christakis, Contractor, Fowler, and Gutmann}{Lazer
  \textit{et~al.}}{2009}]{Lazer:2009aa}
Lazer, D., Pentland, A., Adamic, L., Aral, S., Barab{\'a}si, A.-L., Brewer, D.,
  Christakis, N., Contractor, N., Fowler, J., and Gutmann, M. (2009).
\newblock Computational social science.
\newblock {\em Science\/}~{\em 323\/}(5915), 721--723.

\bibitem[\protect\citeauthoryear{Livingston, Bailey, and Boididou}{Livingston
  \textit{et~al.}}{2018}]{Livingston:2018aa}
Livingston, M., Bailey, N., and Boididou, C. (2018).
\newblock Private sector rents in uk cities: analysis of zoopla rental listings
  data.
\newblock Report, University of Glasgow.

\bibitem[\protect\citeauthoryear{Miller}{Miller}{2010}]{Miller:2010aa}
Miller, H.~J. (2010).
\newblock The data avalanche is here. shouldn't we be digging?
\newblock {\em Journal of Regional Science\/}~{\em 50\/}(1), 181--201.

\bibitem[\protect\citeauthoryear{{Ministry of Housing Communities and Local
  Government (MHCLG)}}{{Ministry of Housing Communities and Local Government
  (MHCLG)}}{2019}]{Housing-Communities:2019aa}
{Ministry of Housing Communities and Local Government (MHCLG)} (2019).
\newblock English housing survey: private rented sector 2016 to 2017.
\newblock Report.

\bibitem[\protect\citeauthoryear{National~Statistician}{National~Statistician}{2012}]{National-Statistician:2012aa}
National~Statistician, . (2012).
\newblock National statistician's review of official housing market statistics.
\newblock Report.

\bibitem[\protect\citeauthoryear{Office~for Statistics~Regulation}{Office~for
  Statistics~Regulation}{2017}]{Office-for-Statistics-Regulation:2017aa}
Office~for Statistics~Regulation, . (2017).
\newblock Statistics on housing and planning in the uk: systematic review of
  public value.
\newblock Report.

\bibitem[\protect\citeauthoryear{Reeves, Clair, McKee, and Stuckler}{Reeves
  \textit{et~al.}}{2016}]{Reeves:2016aa}
Reeves, A., Clair, A., McKee, M., and Stuckler, D. (2016).
\newblock Reductions in the united kingdom's government housing benefit and
  symptoms of depression in low-income households.
\newblock {\em American journal of epidemiology\/}~{\em 184\/}(6), 421--429.

\bibitem[\protect\citeauthoryear{Rugg and Rhodes}{Rugg and
  Rhodes}{2018}]{Rugg:2018aa}
Rugg, J.~J. and Rhodes, D.~J. (2018).
\newblock The evolving private rented sector: Its contribution and potential.

\bibitem[\protect\citeauthoryear{Schwarz et~al.}{Schwarz
  et~al.}{1978}]{schwarz1978annals}
Schwarz, G. et~al. (1978).
\newblock Estimating the dimension of a model.
\newblock {\em The annals of statistics\/}~{\em 6\/}(2), 461--464.

\bibitem[\protect\citeauthoryear{Scottish~Government}{Scottish~Government}{2018}]{Scottish-Government:2018aa}
Scottish~Government, . (2018).
\newblock Scottish household survey 2017: key findings.
\newblock Report.

\bibitem[\protect\citeauthoryear{Thakuriah, Tilahun, and Zellner}{Thakuriah
  \textit{et~al.}}{2017}]{Thakuriah:2017aa}
Thakuriah, P.~V., Tilahun, N.~Y., and Zellner, M. (2017).
\newblock {\em Big data and urban informatics: innovations and challenges to
  urban planning and knowledge discovery}, pp.\  11--45.
\newblock Springer.

\bibitem[\protect\citeauthoryear{Wood}{Wood}{2006}]{wood-2006-book}
Wood, S. (2006).
\newblock {\em Generalized Additive Models: an introduction with R}.
\newblock CRC press.

\end{thebibliography}

\end{document}